\documentclass{aa}  
\usepackage{rotating}
\usepackage{makecell, tabularx}
\usepackage[normalem]{ulem}
\usepackage{multirow}
\usepackage{graphicx}
\usepackage{txfonts}
\usepackage[]{hyperref}
\hypersetup{unicode=true, colorlinks=true, linkcolor=[rgb]{0.53, 0.15, 0.34}, citecolor=[rgb]{0.0, 0.27, 0.42}, filecolor=[rgb]{1.0, 0.13, 0.32}, urlcolor=[rgb]{0.53, 0.15, 0.34}}

\newcommand{\nicer}{{\rm NICER}\xspace}
\newcommand{\cyg}{{\rm Cygnus~X-1}\xspace}

\usepackage[export]{adjustbox}

\graphicspath{{figs/}} 

\begin{document}

   \title{A hidden quasi-periodic oscillation in Cygnus~X-1\\revealed by NICER}

   \subtitle{}

   \author{
   Federico A. Fogantini\inst{1}\thanks{fafogantini@iar.unlp.edu.ar}
    \and
    Federico García\inst{1}
    \and
    Mariano Méndez\inst{2}
    \and 
    Ole K\"{o}nig\inst{3,4}
    \and
    Joern Wilms\inst{4}
    }

   \institute{Instituto Argentino de Radioastronom\'ia (CCT La Plata, CONICET; CICPBA; UNLP), C.C.5, (1894) Villa Elisa, Argentina 
   \and
   Kapteyn Astronomical Institute, University of Groningen, P.O. Box 800, 9700 AV Groningen, The Netherlands 
   \and
   Center for Astrophysics | Harvard \& Smithsonian, 60 Garden Street, Cambridge, MA 02138, USA
   \and
   Dr. Karl Remeis-Observatory and Erlangen Centre for Astroparticle Physics, Universit{\"a}t Erlangen-Nürnberg, Sternwartstr. 7, 96049 Bamberg, Germany
   }

   \date{}

  \abstract
{\cyg is a high-mass black-hole binary system that has been extensively studied across multiple wavelengths since its discovery in 1964. Its rapid temporal and spectral variability in X-rays offer critical insights into the physics of accretion  and the dynamics around black-hole systems. The power spectra of \cyg are generally featureless, and is often modelled with two broad Lorentzian functions, without the need for narrow quasi-periodic oscillations, which are prevalent in other black-hole X-ray binaries.}
{We explore this in light of the recent proposal that some variability components, that are not detected in the power spectra, may be significantly detected in the imaginary part of the cross spectra between two different energy bands and the coherence function. Specifically, we study the power, cross, lag spectra and the coherence function of all available observations of \cyg from the NICER mission up to Cycle 6 looking for these so-called imaginary components.}
{We fitted simultaneously the power spectra of the source in two energy bands, 0.3--2~keV and 2--12~keV, and the real and imaginary parts of the cross-spectrum between the same energy bands, with a multi-Lorentzian model. Under the assumption that each Lorentzian is coherent between the two energy bands, while the Lorentzians are incoherent with one another, our fits predict the intrinsic coherence and phase lags.}
{The intrinsic coherence shows a narrow dip at a frequency that increases from ${\sim}1$~Hz to ${\sim}6$~Hz as the power-law index of the Comptonized component increases from ${\sim}1.8$ to ${\sim}2.4$. Simultaneously, the phase lags show a sudden and steep increase (hereafter referred to as the cliff) at the same frequencies. The dip and the cliff disappear if we use energy bands similar to those of missions like RXTE (e.g., 3--5~keV and 5--12~keV) to compute the coherence and phase-lag spectrum.
A narrow Lorentzian component with a low fractional root mean square amplitude (rms) and a large phase lag is required to effectively reproduce the drop of the intrinsic coherence. The rms and phase-lag spectra of this component change in a systematic way as the source moves in the hardness-intensity diagram.}
{This component, referred to as the imaginary QPO, exhibits behaviour consistent with the canonical type-C QPO, despite being undetectable in the power spectra alone.
Comparison with a similar QPO found in MAXI J1348--630 and MAXI J1820+070 further supports this identification. If our interpretation is correct, this would be the first time that the type-C QPO is detected in \cyg. }


   \maketitle
%

\section{Introduction}
\label{sec:intro}

Accreting black hole binaries (BHBs) exhibit rich phenomenology in their X-ray light curves, with variability occurring over a wide range of timescales, from milliseconds to years (see e.g. the reviews by \citealt{Belloni-2014, Belloni-2016, Ingram-2019}). These sources display different spectral states, namely the soft state, where the X-ray emission is dominated by thermal radiation from the accretion disc \citep{Shakura-1973}, and the hard state, where inverse Compton scattering of soft photons by a hot plasma, or corona, dominates the energy spectrum \citep{Sunyaev-1980, Wilms-2000}. Intermediate states, known as hard- and soft-intermediate states, show characteristics of both regimes \citep{Belloni-2011}. Each state is characterised by unique timing and spectral properties, with radio emission detected in the hard, hard- and soft-intermediate states, but suppressed in the high-soft state \citep{Fender-2009}. 
Note that for historic reasons the ``soft state'' in \cyg is not the same as the high-soft state in transient low mass X-ray binaries. For a comparison between the Hardness-Intensity diagram of \cyg and other transient systems, see Figure~1 of \cite{Konig-2024}.

In a hardness-intensity diagram (HID), transient BHBs follow a q-shaped trajectory that maps out these different states, with transitions between states often accompanied by significant changes in jet activity. Specifically, the transition from the hard-intermediate to the soft-intermediate state, referred to as the jet-line, is associated with discrete radio ejection events \citep{Fender-2009}. This q-track behaviour is seen in other accreting systems, such as neutron star (NS) X-ray binaries, suggesting that fundamental accretion physics is at work in a wide range of objects \citep{Motta-2017}.

\cyg, one of the most studied black hole binaries, is a persistent high-mass X-ray binary (HMXB) that accretes from the stellar wind of the supergiant HDE 226868 \citep{Walborn1973}. Hosting a 21.2$\pm$2.2 M$_\odot$ black hole \citep{Miller-Jones-2021}, \cyg is key to understanding accretion physics and jet launching processes. It exhibits frequent state transitions, sometimes very rapid, crossing the jet line in the intermediate states \citep{Pottschmidt-2003}. The system is characterised by a relatively small bolometric luminosity variation of a factor of ${\sim}$3-4 between states, and it never fully enters the disc-dominated regime \citep{Wilms-2006}. Despite being constrained to the lower branch of the q-track \citep{Konig-2024}, \cyg shows rich multi-wavelength variability, including a radio jet during its hard state and polarised emission at high energies \citep{Krawczynski2022, Chattopadhyay-2023, Jana-2023}.

The timing properties of BHBs can be investigated through Fourier analysis, where power density spectra (PDS) provide insights into the variability components \citep{vanderKlis-1989}. These PDS reveal broadband variability extending up to 100~Hz, along with narrow peaks corresponding to quasi-periodic oscillations (QPOs; \citealt{Nowak-2000, Belloni2002}). Low-frequency QPOs (LFQPOs), with frequencies from a few mHz to ${\sim}$30 Hz, are commonly observed in BHBs and are classified into three types: A, B, and C, based on their quality factor\footnote{The quality factor $Q$ of a Lorentzian profile is defined as $\nu_0/\Delta$, where $\nu_0$ is the centroid frequency and $\Delta$ is the full-width at half maximum.}, fractional rms amplitude, phase lags, and the strength of the underlying noise \citep{Casella-2005}. High-frequency QPOs, up to ${\sim}$350 Hz, are rarer, but their presence and characteristics suggest a link to similar oscillations seen in neutron star systems, implying a shared origin \citep{Titarchuk1998, Belloni-2012,Mendez-2013}.

The rms amplitude of PDS components, particularly the $0.1-10$~Hz integrated rms amplitude, serves as a tracer of accretion regimes in BHBs \citep{Belloni-2011}. The rms amplitude of QPOs tends to increase with energy, flattening above 10~keV, either increasing or decreasing at higher energies \citep{Yan-2018, Zhang-2020,Yang2024,Zhu2024}. On the other hand, phase lags, which describe the delay between 
correlated light curves in two different energy bands, provide vital information about the variability components in the PDS at specific frequencies \citep{Belloni2002,MDarias2011}. These lags can increase, decrease, or remain constant with energy, depending on the system and the characteristic frequency of the variability components \citep{Reig-2000, Zhang-2020}. 

In the case of \cyg, the hard state PDS is commonly modeled using a combination of Lorentzian components \citep{Nowak-2000}, with rms variability reaching $30-40$\% in the $2-13$ keV range \citep{Pottschmidt-2003}. As the source transitions into the soft state, the rms drops to $10-20$\% \citep{Grinberg-2014, Konig-2024}, and the PDS becomes dominated by red noise, i.e., a power law. Time-lags measurements in \cyg have revealed complex behaviour, whose interpretation is further complicated by the possibility of contributions from processes such as reprocessing in the accretion disc or scattering in the stellar wind \citep{Grinberg-2014, Lai-2022, Harer-2023, Konig-2024}.

Additionally, the coherence function, which measures the correlation between variability in different energy bands \citep{Vaughan-1997}, offers further insight into the accretion dynamics. In some systems, such as MAXI J1348-630, MAXI J1820+070 and \cyg, the coherence drops at specific frequencies, in hand with a sudden increase of the phase lags \citep{Nowak1999, Ji-2003, Konig-2024,Alabarta2025}. 
It has been proposed \citep{Konig-2024} that the drop of the intrinsic coherence could be due to the ``beat'' of two or more dominant variability components. The drop would then happen at the frequency difference between the two variability components, which can then produce a modulation in the observed variability when the components are sufficiently coupled.
Another way the intrinsic coherence can drop at a specific frequency is if two or more components with different amplitudes and phases of their cross-vectors contribute to the variability over the same frequency range \citep{Vaughan-1997, Mendez-2023}.

In this study, we leverage the high throughput and precise timing capabilities of NICER to explore the broadband variability of \cyg across its full range of observed spectral states, utilising data from several observing cycles. The structure of this paper is as follows: In Sect. 2 we present the NICER observations and data reduction processes. In Sect. 3 we outline the timing analysis, emphasising the variability components across distinct accretion states. Finally, in Sect. 4, we discuss our findings on \cyg and compare them with other sources that have shown similar behaviours.


\section{Data Analysis}
\label{sec:data}

The Neutron star Interior Composition ExploreR (\nicer; \citealt{Gendreau-2016}) is an instrument on board the International Space Station. NICER's X-ray Timing Instrument (XTI) detector is sensitive to X-rays in the $0.3-12$ keV energy range with a temporal resolution of $40$~ns. 
We analysed all available \nicer archival observations of \cyg up to Cycle 6, which consist of more than 100 pointings performed between June 2017 and June 2023, with exposures ranging from ${\sim}0.1$ to ${\sim}30$~ks. 
We used the \nicer reduction tool {\tt nicerl2} from the {\tt HEASoft}~v.6.33.1 package with CALDB version {\tt xti20240206} to process the data. To filter out events that could contaminate the light curves, we selected events with undershoot rate less than 200~counts~s$^{-1}$, overshoot rate less than 5~counts~s$^{-1}$, and cutoff rigidity (COR\_SAX) greater than 1.5~GeV~counts$^{-1}$. 
Energy spectra were extracted using {\tt nicerl3-spect} routine with the Space Weather background model and a minimum of 30 counts per bin. We used XSPEC~v12.14.0h \citep{xspec} to model each spectrum and extracted fluxes using the {\sc cflux} convolution model.

We computed PDS from cleaned event files with the {\tt GHATS} software\footnote{\url{http://www.brera.inaf.it/utenti/belloni/GHATS_Package/Home.html}}, using time segments of 128~s and a Nyquist frequency of 1024~Hz. We subtracted the contribution of the Poisson noise from the PDS \citep{Zhang-1995}, which we estimated from the average of powers above 500~Hz, and normalised the PDS to units of rms$^2$ Hz$^{-1}$. When converting the PDS to rms units, we do not apply a background correction since, compared to the source, the background count rate is negligible in all observations.

We defined three energy bands for the extraction of light curves and Fourier products: 0.3--12~keV (total), 0.3--2~keV (soft) and 2--12~keV (hard). 
Real and imaginary parts of the cross spectra (CS), phase lags and intrinsic coherence were extracted using the formulae described in \cite{Mendez-2023}. 
To account for the partial correlation of photons simultaneously present in the narrow and total energy bands, we subtract the average of the real part of the CS calculated over a frequency range where the source has no contribution.
We applied a logarithmic rebin to all Fourier products such that the size of each frequency bin increases by $10^{1/100}$ with respect to the previous one.
Finally, we fitted simultaneously the soft and hard PDS, as well as the rotated\footnote{As described in \cite{Mendez-2023}, $\pi/4$ radians are added to the argument of both components of the CS, which is appropriately accounted for afterwards when modeling.} real and imaginary parts of the CS using XSPEC~v.12.14.0h \citep{xspec} from ${\sim}0.004$~Hz up to $200$~Hz. 

In \hyperref[tab:data]{Table~\ref{tab:data}} we show the complete NICER dataset used in this work. The reported exposures times correspond to the sum of the total non-zero GTIs created by GHATS after processing the cleaned event files.


\begin{figure}
    \centering
    \includegraphics[width=0.5\textwidth,keepaspectratio]{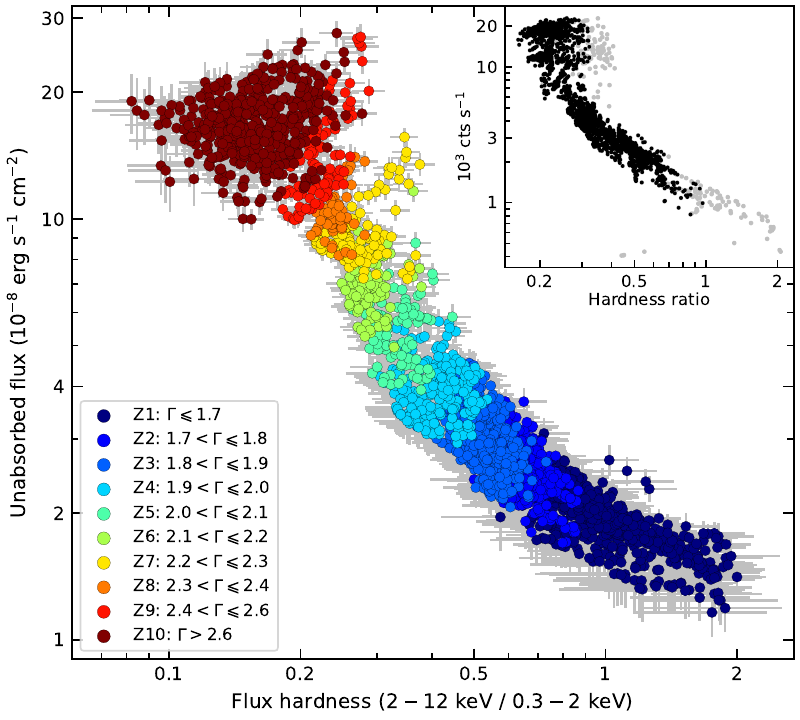}
    \caption{Hardness-Intensity Diagram (HID) computed from the unabsorbed fluxes of the NICER dataset of \cyg. The hardness ratio in $x$ axis is defined as the ratio of the unabsorbed flux in the $2-12$ keV band to that in $0.3-2$ keV band, while the $y$ axis is the unabsorbed flux in the $0.3-12$ keV band. The fluxes were obtained from the fitting each 128~s segment with a model consisting of disc blackbody plus a power-law component, both affected by interstellar absorption. The different colours indicate the 10 distinct regions we use in this paper, parametrized by the increasing spectral index, $\Gamma$, of the power-law component (see legend). The grey error bars in the main panel indicate the $1\sigma$ errors of the fluxes and hardness ratios. Top-right inset panel: HID of \cyg computed from the observed count rates in the same energy bands as in the main panel; the black points correspond to the segments we used in this work, whereas the grey points are the segments that we discarded in our analysis (see text for details)}
    \label{fig:hid}
\end{figure}

\section{Results}
\label{sec:results}

\subsection{Constructing an absorption corrected HID}

To analyse the entire NICER dataset, we first construct the HID using unabsorbed fluxes for the same energy bands and time segments used to create the power spectra. For this, we fit the energy spectrum of each 128~s segment in the 0.3--12~keV energy range with the model {\sc tbabs$\times$(diskbb+powerlaw)}, to account for the thermal emission of the accretion disc and the Comptonized emission of the corona, both affected by interstellar absorption. We used a power law instead of a Comptonization model to facilitate the comparison with previous studies (see e.g., \citealt{Done1992,DiSalvo2001,Zdziarski2002,Wilms-2006}).
For the {\sc tbabs} component we used the cross-sections tables of \cite{Verner-1996} and the abundances of \cite{Wilms-2000}.
We applied a 5\% systematic error, higher than the recommended value, to account for our use of a simplified spectral model (a disc plus a Comptonizing corona) that excludes broad emission lines, relativistic reflection, and non-solar ISM abundances. Following the approach of \cite{Konig-2024}, this method enables the construction of a hardness-intensity diagram based on intrinsic source fluxes, minimizing the impact of interstellar absorption.

We excluded segments that yielded reduced $\chi^2$ greater than 2, which mostly correspond to intervals of high absorption ($N_{\rm H}>0.7\times10^{22}$~cm$^{-2}$), low/high disc temperatures ($kT_{\rm dbb}<0.1$ keV or $kT_{\rm dbb}>0.5$ keV) and low/high spectral index ($\Gamma<1.4$ or $\Gamma>3.4$). These excluded segments are most likely affected by the stellar wind of the companion \citep{Wilms-2006}.
After all this filtering, we kept 92.7\% of the original set of segments of 128~s, corresponding to a total of ${\sim}380$~ks of NICER data across the entire HID, as displayed in \hyperref[fig:hid]{Fig.~\ref{fig:hid}}. 

For comparison, we include the instrument-dependent HID on the same Figure, where we distinguish the time segments that we retained (black) from the segments that we discarded (gray). We note that the hardness ratio in both cases spans about a decade, while the 0.3--12~keV intensity spans almost two orders of magnitude, whereas the unabsorbed 0.3--12~keV flux changes by a factor of ${\sim}30$.

\begin{figure*}
    \centering
    \includegraphics[width=0.99\textwidth,keepaspectratio]{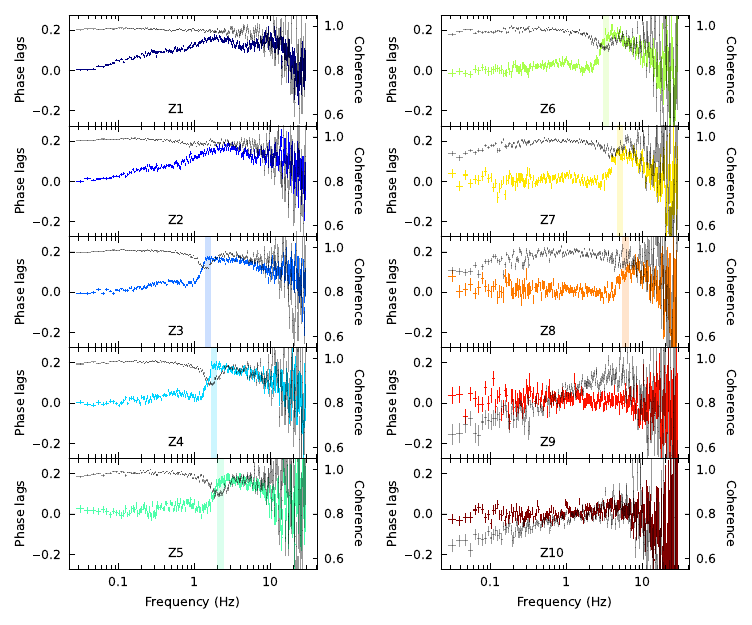}
    \caption{Phase lags in radians (colour points) and intrinsic coherence (gray points) of each region of the HID of \cyg, restricted to frequencies between 0.02 Hz and 30 Hz. The vertical stripe indicates the minimum of the coherence drop and the maximum of the phase-lags cliff. Phase lags and coherence is computed between 0.3-2~keV and 2--12~keV energy bands, using the latter as reference band. }
    \label{fig:plags}
\end{figure*}

We segmented the HID by the power-law spectral index in bins of width $\Delta\Gamma=0.1$, starting at the low-hard state ($\Gamma\simeq1.4$), and finishing at the softest state registered ($\Gamma\simeq3.4$). After visually inspecting and comparing the frequency-dependent phase-lags and intrinsic coherence of each bin, we combined the data to form 10 distinct regions, $Z1$ to $Z10$, across the HID. These regions are colour coded in \hyperref[fig:hid]{Fig.~\ref{fig:hid}}, where we indicate the corresponding values of the spectral index. Using the same colour scheme, we present in \hyperref[fig:plags]{Fig.~\ref{fig:plags}} the phase lags and coherence function between the 0.3--2~keV and 2--12~keV energy band of each region. 

\renewcommand{\arraystretch}{1.25}
\begin{table}[]
    \centering
    \caption{\label{tab:zoneinfo} Properties of the defined HID regions.}
    \resizebox{0.99\columnwidth}{!}{
    \begin{tabular}{c | c c c c }
    Region & Exposure (ks) & Segments & $kT_{\rm dbb}$ (keV) & $\Gamma$\\ \hline
    Z1  & 89.3 & 698 & $0.22\pm0.02$ & $<1.7$ \\
    Z2  & 54.3 & 424 & $0.23\pm0.01$ & $1.7-1.8$ \\
    Z3  & 75.0 & 586 & $0.25\pm0.01$ & $1.8-1.9$ \\
    Z4  & 33.0 & 258 & $0.25\pm0.02$ & $1.9-2.0$ \\
    Z5  & 12.8 & 100 & $0.26\pm0.02$ & $2.0-2.1$ \\
    Z6  & 14.5 & 113 & $0.29\pm0.02$ & $2.1-2.2$ \\
    Z7  & 15.9 & 124 & $0.32\pm0.01$ & $2.2-2.3$ \\
    Z8  & 8.6  & 67 & $0.34\pm0.01$ & $2.3-2.4$ \\
    Z9  & 12.4 & 97 & $0.38\pm0.02$ & $2.4-2.6$ \\
    Z10 & 65.9 & 515 & $0.42\pm0.02$ & $>2.6$ \\ \hline
    \end{tabular} 
    }
    \tablefoot{Summary of the properties of each identified region of the HID of \cyg: total accumulated exposure time (ks), total number of accumulated 128~s segments, average and standard deviation of the disc-blackbody temperature and power-law index intervals.}
\end{table}

The most prominent feature seen in \hyperref[fig:plags]{Fig.~\ref{fig:plags}} (and the main focus of this work) is the appearance of a sudden and steep increase in phase lags, hereafter named the cliff, in conjunction with a sudden drop in the coherence (see \citealt{Konig-2024}).\footnote{From now on, cliff refers to the phase lags feature, while drop refers to the relative narrow feature in the coherence function.} 
These two features appear to correlate with frequency and spectral hardness, as the frequency at which they appear shifts from ${\sim}1.5$ Hz in $Z3$ ($\Gamma\simeq1.8$, the first HID region where it appears) up to ${\sim}6$ Hz in $Z8$ ($\Gamma\simeq2.4$). In \hyperref[fig:plags]{Fig.~\ref{fig:plags}} we mark the centroid frequency of the coherence drop using a vertical stripe, which was derived from modelling the Fourier products, as we describe in \hyperref[sec:plags]{Sect.~\ref{sec:plags}}.

To identify more precisely the value of the photon index at which the cliff/drop features appear, we segmented the HID around $Z2$ and $Z3$ by spectral index in bins of size $\Delta\Gamma=0.05$. We find that the feature appears rather abruptly at $\Gamma\simeq1.8$, at a frequency of approximately $1$~Hz.

\begin{figure}[h!]
    \centering
    \includegraphics[width=0.99\columnwidth]{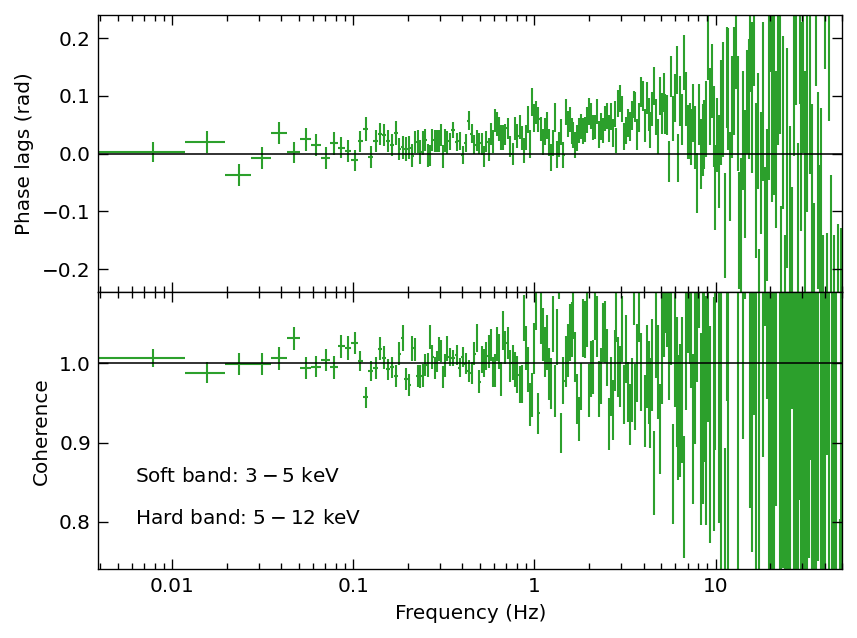}
    \caption{Phase lags and intrinsic coherence function of region Z3 of the HID of \cyg, computed between the 3-5~keV and 5--12~keV energy bands. No evidence of the coherence drop or the phase-lags cliff at ${\sim}$1.5~Hz is apparent (see \hyperref[fig:plags]{Fig.~\ref{fig:plags}}). The horizontal lines in each panel indicate the reference levels for zero phase lag and unity coherence respectively.}
    \label{fig:highecut}
\end{figure}

\subsection{Dependence of the coherence drop upon energy}

\cyg has been observed repeatedly since its discovery in 1964. For instance, RXTE has accumulated close to 4.8~Ms in exposure time across several observing campaigns (e.g., \citealt{Belloni1996,Revnivtsev2000,Grinberg-2014}). However, the drop and cliff features presented here have not been reported before up to the appearance of NICER. As noted by \cite{Konig-2024}, the key difference were the data at energies below 3 keV, which RXTE could not cover.

To illustrate this, in \hyperref[fig:highecut]{Fig.~\ref{fig:highecut}} we show the phase lags and intrinsic coherence of region $Z3$ but computed from energy bands similar to those commonly used with RXTE data: 3--5~keV and 5--12~keV. It becomes apparent from this Figure that neither the cliff feature in the phase lags, nor the drop feature in the coherence are present. As shown in Figure~8 of \cite{Konig-2024}, the feature in the coherence function appeared only when they considered a band below ${\sim}$1.5--2~keV.

\subsection{Constant phase-lag model}
\label{sec:plags}

\begin{figure} 
\centering
\includegraphics[width=0.49\textwidth,keepaspectratio]{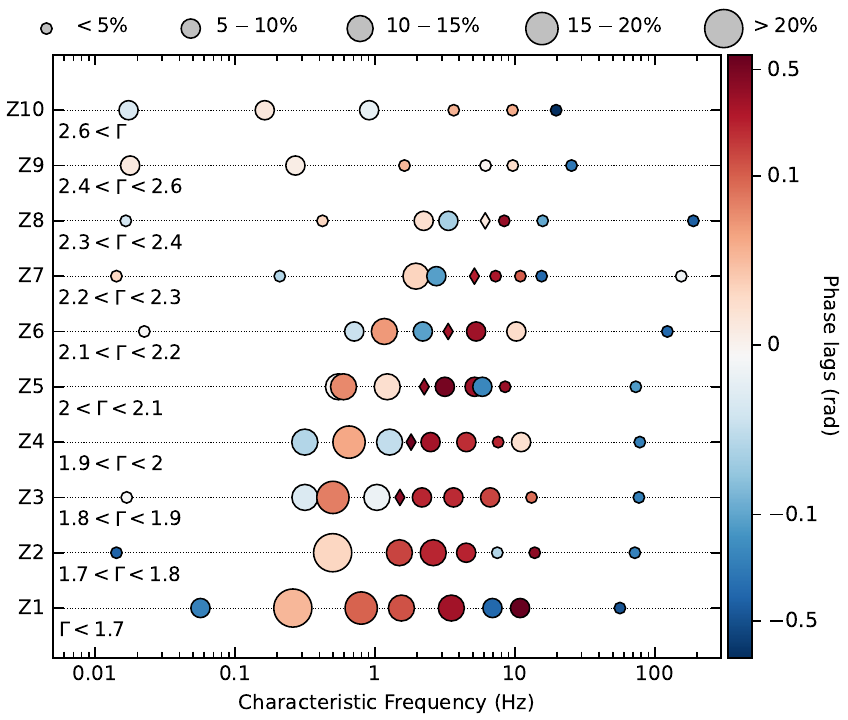}
\caption{Bubble plot of every Lorentzian in each HID region (vertical axis) of \cyg, as a function of their characteristic frequency (horizontal axis).
The area of each bubble is proportional to the covariance rms amplitude of the Lorentzian. The colour scheme indicates the corresponding phase lag.
The diamond marker highlights the narrow QPO responsible for the sharp coherence drop, which becomes apparent in Zones 3 to 8, corresponding to intermediate states with $1.8 < \Gamma < 2.4$.}
\label{fig:pbubbles}
\end{figure}

Following \cite{Mendez-2023}, we used {\sc XSPEC} to fit simultaneously the PDS in two energy bands (0.3--2~keV and 2--12~keV), as well as the real and imaginary parts of the cross spectrum in the same two bands. 
At all Fourier frequencies, the magnitude of the real part of the cross-spectrum is consistently larger than the magnitude of the imaginary part. To improve the stability of fitting procedures, such as the Levenberg–Marquardt algorithm in {\sc XSPEC}, which work more effectively when parameters are of similar magnitude, we rotated the cross vectors by 45 degrees.~\footnote{See Section 15.5.2 of \cite{Press2007} for a discussion on the importance of the correct scaling of the Hessian matrix in the context of the Levenberg–Marquardt algorithm.}
This rotation equalizes the real and imaginary components for cross vectors with zero phase lag, without altering the modulus or the fit parameters. In the following Figures, we present the rotated components $({\mathcal R}\cdot \cos{\pi/4} - {\mathcal I}\cdot \sin{\pi/4}) \propto {\mathcal R-I}$ and $({\mathcal R}\cdot \sin{\pi/4} + {\mathcal I}\cdot \cos{\pi/4}) \propto {\mathcal R+I}$, which remain positive, enabling the use of logarithmic axes. Phase lags are reported after subtracting $\pi/4$, returning the values to the reference frame of the original non-rotated cross vector.

The PDS in the two bands are modelled using Lorentzian functions, with the centroid frequency and width of each Lorentzian tied across all Fourier products. To model the CS components, we apply two different variants: a constant phase-lag and a constant time-lag model. Each model assumes that the phase lags of each Lorentzian component, either independent of frequency (constant phase lag model, hereafter $\phi$-model) or linearly dependent upon frequency (constant time lag model, hereafter $\tau$-model). 
To keep the analysis as straightforward as possible, we adopt the $\phi$-model in the main text. However, we also fit the data using the $\tau$-model, and the corresponding results are presented in \hyperref[app:taumodel]{Appendix~\ref{app:taumodel}}.

Each Lorentzian component in the cross spectrum is then multiplied by the cosine (real part), or sine (imaginary part) of the phase-lags model.
Thus, each Lorentzian model has a total of 6 free parameters across all four Fourier components: centroid frequency and width, soft and hard PDS normalizations, phase lag and CS normalization. 

 Each region of the HID is described with a variable number of Lorentzians. We consider that a Lorentzian component is significant if at least one of the three free normalizations is not consistent with zero at the 3$\sigma$ level.  As noted by \cite{Mendez-2023}, if each Lorentzian is perfectly coherent in the two energy bands, the normalization $C$ of each Lorentzian in the CS can be written in terms of the normalizations of the soft ($N_s$) and hard ($N_h$) PDS normalizations as $C=(N_s N_h)^{1/2}$. Here, we did not link these parameters during the fit, but checked that after the fit, this relationship held for all components within uncertainties.

The best fitting parameters as well as the $1\sigma$ uncertainties were derived from MCMC simulations using XSPEC {\sc chain} command. We verified that each chain converged correctly by visually inspecting each parameter trace plot, and by simultaneously verifying that the autocorrelation time is close to unity\footnote{\url{https://emcee.readthedocs.io/en/stable/tutorials/autocorr/}}. 
To achieve this, we needed chains of length 10$^6$, burning the first $5\times10^5$ steps. In \hyperref[tab:plorstats]{Table~\ref{tab:plorstats}} and \hyperref[tab:tlorstats]{Table~\ref{tab:tlorstats}} we present a summary of the entire set of properties of the Lorentzian components used to model the 10 regions of the HID with, respectively, the constant phase-lag and the constant time-lag model.

\begin{figure*} 
\centering
\includegraphics[width=0.99\textwidth,keepaspectratio]{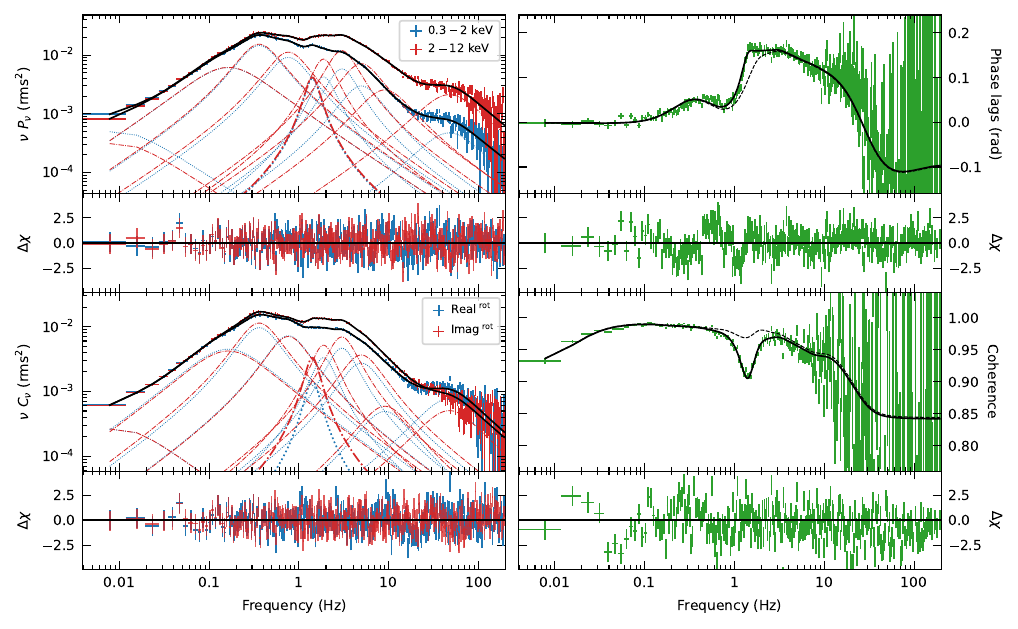}
\caption{Constant phase-lag model applied to region $Z3$ of the HID of \cyg. 
The top left panels show the soft ($0.3-2$~keV) and hard ($2-12$~keV) PDS  in rms$^2$ units and residuals. 
The bottom left panels show the real and imaginary parts of the cross-spectrum (with the cross vector rotated by $\pi/4$) in rms$^2$ units and the residuals. 
The top right panels show the phase-lags (rad) with the derived model and residuals. 
The bottom right panels show the intrinsic coherence with the derived model and residuals. 
As described in the text, the model lines drawn in the right panels have not been fitted to the phase-lags and the coherence, but have been derived from the fits to the PDS and CS in the left panels.
The QPO responsible for the coherence drop is highlighted over the other Lorentzians using a thicker line (at $\nu \sim 1.3$~Hz).
The solid lines in the plots of the phase lags and coherence function show the derived model of those two quantities for 10 Lorentzians (see text). Because the total phase lag spectrum and coherence function are not a combination of additive components, in these plots we cannot show the contribution of each Lorentzian. To show the effect of the imaginary QPO, the dashed lines in the plots show the derived models without that Lorentzian (without refitting the data).}
\label{fig:plors}
\end{figure*}

\begin{table}[]
    \centering
    \caption{\label{tab:stats} Best fit statistics summary.}
    \resizebox{0.99\columnwidth}{!}{
    \begin{tabular}{c|c c|c c}
    & \multicolumn{2}{c|}{Constant phase-lag model} & \multicolumn{2}{c}{Constant time-lag model} \\
    Region & $\chi^2$ / dof & Lorentzians & $\chi^2$/dof & Lorentzians \\ \hline 
    1 & 1245.2 / 1152 & 8 & 1279.1 / 1146 & 9 \\
    2 & 1328.6 / 1152 & 8 & 1370.6 / 1152 & 8 \\
    3 & 1148.8 / 1140 & 10 & 1137.5 / 1140 & 10 \\
    4 & 1183.0 / 1146 & 9 & 1171.8 / 1146 & 9 \\
    5 & 1169.0 / 1152 & 8 & 1115.7 / 1146 & 9 \\
    6 & 1286.5 / 1152 & 8 & 1161.8 / 1140 & 10 \\
    7 & 1279.6 / 1146 & 9 & 1264.0 / 1146 & 9 \\
    8 & 1396.9 / 1152 & 8 & 1368.7 / 1152 & 8 \\
    9 & 1201.0 / 1164 & 6 & 1227.3 / 1170 & 5 \\
    10 & 1386.8 / 1164 & 6 & 1399.3 / 1170 & 5 \\
    \hline
    \end{tabular} 
    }
    \tablefoot{Summary of the best-fit statistics for each region of the HID of \cyg, comparing the constant phase-lag and the constant time-lag models. The fit statistics are given by the total $\chi^2$ and number of degrees of freedom (dof). The number of Lorentzian components used is shown for both the phase-lag and time-lag models.}
\end{table}


To better visualise the entire collection of Lorentzian components and their respective properties shown on \hyperref[tab:plorstats]{Table~\ref{tab:plorstats}}, we constructed bubble plots. \hyperref[fig:pbubbles]{Fig.~\ref{fig:pbubbles}} displays the characteristic frequency of each Lorentzian (horizontal axis), defined as $(\nu_0^2+\Delta^2)^{1/2}$, where $\nu_0$ and $\Delta$ are the centroid frequency and FWHM of the corresponding Lorentzian \citep{Belloni-2022}. On the vertical axis we display each HID region (increasing upwards). The bubble area is proportional to the rms amplitude of the cross-vector, also known as the covariance amplitude. A colour map is used to display phase lags (in radians). In \hyperref[tab:stats]{Table~\ref{tab:stats}} we show the total $\chi^2$ and degrees of freedom corresponding to each HID region and applied model, as well as the number of Lorentzians used per region and model.

The entire collection of soft and hard PDS, as well as the rotated real and imaginary parts of the CS of \cyg, between ${\sim}0.004$ and $200$ Hz (${\sim}5$ decades in frequency), can be generally described with ${\sim}6$ broad Lorentzians with quality factor $Q\lesssim0.5$, with relative contributions to the total rms that vary according to the spectral hardness of the source. This description of ${\sim}6$ broad components is most notably seen in the hardest, low-flux observations ($\Gamma<1.8$) and the softest, high-flux observations ($\Gamma>2.4$). In the observations at intermediate hardness ($1.8<\Gamma<2.4$), we notice that to fit simultaneously the PDSs and CSs correctly we need to add extra narrow components ($Q>0.5$), specifically between 1 and 10 Hz.

The necessity to add narrow Lorentzian components to model adequately the Fourier products may arise from the combination of various factors. First and foremost, NICER's high throughput and low energy coverage may reveal components that up to now were undetectable with previous, less sensitive, missions, or missions not covering the same energy band. Moreover, we achieved a great signal to noise ratio in each HID region by averaging several segments (see \hyperref[tab:stats]{Table~\ref{tab:stats}}), and by afterwards logarithmically rebining each Fourier product.
Secondly, the simultaneous PDS+CS modelling utilised here can reveal weak or hidden variability components that may otherwise be lost by fitting only the PDS \citep{Mendez-2023}.

When looking at the evolution of all the Lorentzian components across the HID in \hyperref[fig:pbubbles]{Fig.~\ref{fig:pbubbles}}, a clear shift towards higher frequencies can be seen as the source spectrum softens. 
For example, the Lorentzian that in the hardest region (Z1) starts at $\sim0.3$ Hz with $\gtrsim20\%$ rms, ends up in region Z8 at $\sim2$ Hz with $\lesssim10\%$ rms. 
This is the component at the lowest frequency out of the two broad components used historically to model the PDS in the hard state of \cyg (see e.g., \citealt{Pottschmidt-2003}).
This phenomenon of increasing frequency shift accompanied by a decreasing rms is seen in almost all components in the regions between Z1 and Z8. In regions Z9 and Z10, where $\Gamma>2.4$, the PDS show the typical power-law shape studied extensively in previous campaigns of \cyg \citep{Pottschmidt-2003}.

The PDS and CS of \cyg also show one broad ($Q \lesssim 0.1$) Lorentzian component at very low frequency  that remains between 0.01~Hz and 0.02 Hz across the entire HID, with rms close to $5$\% except for the very soft regions were it reaches ${\sim}10$\%. Only in regions Z4 and Z5 ($1.9<\Gamma<2.1$) was this component not significant enough to fit it properly. 
These components display small phase lags ($|\Delta\phi| \lesssim 0.01$), as shown in white colour in \hyperref[fig:pbubbles]{Fig.~\ref{fig:pbubbles}}. 

Finally, we notice in \hyperref[fig:pbubbles]{Fig.~\ref{fig:pbubbles}} the significant detection of a high frequency (HF) component both in the PDS and CS at frequencies above $50$~Hz in the hardest region ($\Gamma\simeq1.4$), reaching up to ${\sim}150$~Hz in region Z8 ($\Gamma\simeq2.4$). 
This HF component is quite broad ($Q\sim0.5$) and weak (rms${\sim}5$\%), with large negative phase lags (close to $1$~rad). We will further analyse this HF component in \hyperref[sec:bump]{Sect.~\ref{sec:bump}}.

Both these LF and HF components can be seen in \hyperref[fig:plors]{Fig.~\ref{fig:plors}}. On the left side of each plot we show the soft and hard PDS as well as the rotated real and imaginary parts of the CS of region Z3 ($1.8<\Gamma<1.9$). On the right hand side of each plot we show the phase lags and intrinsic coherence together with their respective derived models. This region was successfully modelled using 10 Lorentzians, yielding a $\chi^2$ of $1148.8$ for $1140$ degrees of freedom. The residuals show some structure, but the addition of more Lorentzians components does not improve the fit. The Lorentzian component responsible for the coherence drop seen in this Figure is shown with a thicker line. This highlighted component of region Z3 corresponds to the Lorentzian shown with a diamond marker in \hyperref[fig:pbubbles]{Fig.~\ref{fig:pbubbles}}.


\subsection{Frequency dependence of the intrinsic coherence and phase lags}

In \hyperref[fig:pbubbles]{Fig.~\ref{fig:pbubbles}} the diamond marker shows the Lorentzian component that coincides with, and we attribute to, the coherence drop. 
This component can be traced from region Z3 at ${\sim}$1.5 Hz up to region Z8, reaching ${\sim}$6~Hz. Although \hyperref[fig:plags]{Fig.~\ref{fig:plags}} shows that in region Z2 there is a slight coherence drop at ${\sim}1$ Hz, we could not constrain adequately any narrow component in this region. 
\hyperref[fig:pbubbles]{Fig.~\ref{fig:pbubbles}} also shows that the phase lag of the highlighted component varies between 0.2~rad and 0.5~rad from regions Z3 to Z7, and then decreases below 0.1~rad in region Z8. 
The covariance rms amplitude of that component remains below $5\%$ in all 6 regions where it is detected, with a quality factor above 2 (see \hyperref[tab:plorstats]{Table~\ref{tab:plorstats}} for details).

An example of how this component fits in between stronger and broader ones can be seen in \hyperref[fig:plors]{Fig.~\ref{fig:plors}}. We note that this component (highlighted using a thicker line) is most significantly detected in the imaginary part of the cross spectrum (hence the large phase lag), and the rms amplitudes (proportional to the root square of the normalizations in the PDS) in the soft and hard bands are very similar in magnitude. This behaviour is very different from that of the other Lorentzian components; which all show that the hard dominates over the soft component by factors of a few. 
This is the reason why we will call the Lorentzian component that causes the cliff in the phase-lag spectrum and the drop in the coherence the imaginary QPO. 

We tested the scenario shown in \hyperref[fig:plors]{Fig.~\ref{fig:plors}} without the inclusion of the imaginary QPO. This yielded a $\chi^2$ of $1526.8$ over $1146$ degrees of freedom. For comparison, the best fit for region $Z3$ shown in \hyperref[tab:stats]{Table~\ref{tab:stats}} resulted in a $\chi^2$ of $1148.8$ with $1140$ degrees of freedom. Additionally, the residuals of the $0.3-2$~keV PDS exhibit a $4\sigma$ structured excess around the frequency of the imaginary QPO, at approximately ${\sim}1.5$~Hz. Similarly, the residuals of the imaginary part of the CS shows a $3\sigma$ structured excess at this frequency. Meanwhile, the residuals of the 2--12~keV PDS and the real part of the CS display a structured excess around the same frequency, close to $1\sigma$, though not statistically significant. To show the contribution of the narrow imaginary QPO component to the phase-lag spectrum and the intrinsic coherence function, in the right panels of \hyperref[fig:plors]{Fig.~\ref{fig:plors}} we show with dashed lines the derived models without that component, without refitting the data.


 \subsection{Energy dependence of the amplitude and phase lags of the imaginary QPO}

\begin{figure}
    \centering
    \includegraphics[width=0.99\linewidth]{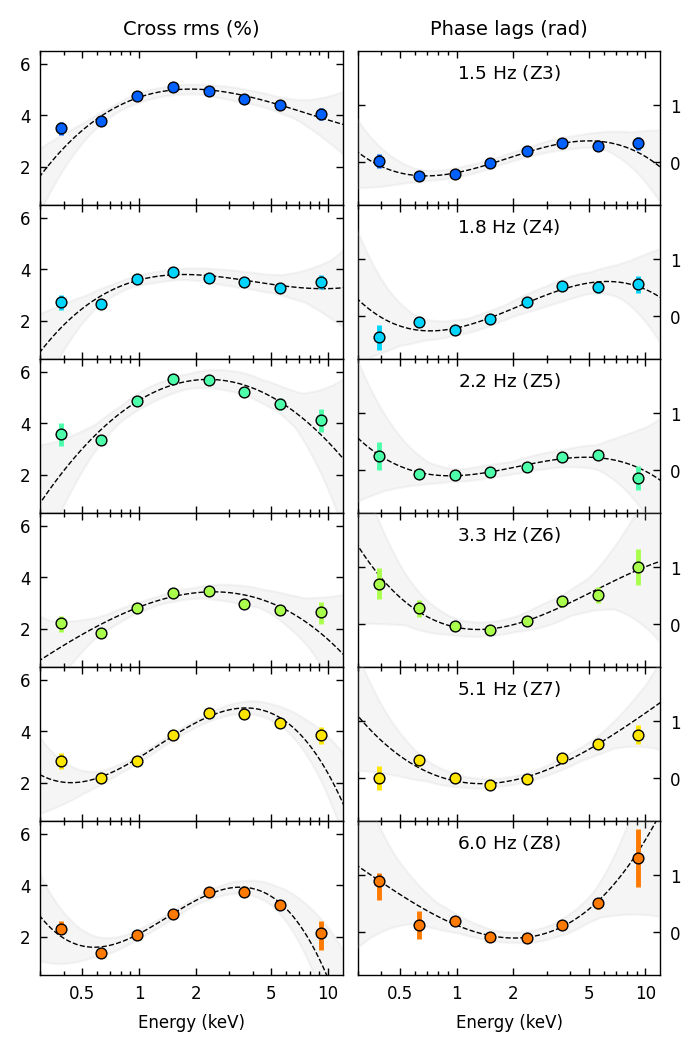}
    \caption{Covariance rms amplitude (\%, left column) and phase lags (radians, right panel) of the imaginary QPO that causes the coherence drop of \cyg as a function of energy. The centroid frequency of the imaginary QPO is indicated in each panel. A cubic polynomial is fitted in each case to track the corresponding minima or maxima. In each panel we show cubic model realisations in grey colour depicting the $1\sigma$ confidence range of the fits.}
    \label{fig:rmslag}
\end{figure}

\begin{table}[]
    \centering
    \caption{\label{tab:minmax} Energies of the rms and phase-lag extrema.}
    \begin{tabular}{c c c c}
        Region & $\nu_{\rm QPO}$ [Hz] &  $E_{\rm max}^{\rm rms}$ [keV] & $E_{\rm min}^{\rm lag}$ [keV] \\
        \hline
         Z3 & 1.5 & $1.9\pm0.1$ & $0.7\pm0.1$ \\
         Z4 & 1.8 & $1.8\pm0.2$ & $0.7\pm0.1$ \\
         Z5 & 2.2 & $2.4\pm0.1$ & $0.9\pm0.1$ \\
         Z6 & 3.3 & $2.5\pm0.2$ & $1.3\pm0.1$ \\
         Z7 & 5.1 & $3.8\pm0.1$ & $1.4\pm0.1$ \\
         Z8 & 6.0 &  $3.5\pm0.1$ & $2.0\pm0.1$ \\
    \end{tabular}
    \tablefoot{Energies of the maximum of the covariance rms amplitude and minimum of the phase lags for each region of the HID of \cyg, where the imaginary QPO ($\nu_{\rm QPO}$) is detected (see \hyperref[fig:rmslag]{Fig.~\ref{fig:rmslag}}). Uncertainties are estimated at $1\sigma$ level.}
\end{table}

To analyse the rms energy spectrum of the imaginary QPO, we constructed PDS of regions 3 to 8 in the following energy bands: 0.3--0.5~keV, 0.5--0.9~keV, 0.9--1.2~keV, 1.2--1.9~keV, 1.9--2.9~keV, 2.9--4.5~keV, 4.5--7~keV and 7--12~keV. We used the full 0.3--12~keV energy band as reference to compute the CS. We then fitted the full band PDS, the band specific PDS, and the real and imaginary parts of the CS of each region using {\sc XSPEC}. 
The complete rms and phase-lag spectrum of the imaginary QPO for each region is presented in \hyperref[fig:rmslag]{Fig.~\ref{fig:rmslag}}.

The rms spectrum of the imaginary QPO shows a wave like structure across all six regions, where one or two local extrema can be identified in the 0.3--12 keV energy range.
The main differences observed between regions are the average rms level and the energy where the rms peaks. There is no clear dependence of the average rms level upon the spectral hardness, but there is an apparent systematic change of the rms peak energy with the spectral hardness. In region Z3 ($\Gamma{\simeq}1.8$), the rms peaks at ${\sim}1.5$ keV, whereas in region Z8 ($\Gamma{\simeq}2.4$) the rms maximum is at about ${\sim}2.5$ keV.

The phase-lag energy spectrum, on the other hand, shows an elongated U shape that evolves as the source moves in the HID. 
Interestingly, as the frequency of the imaginary QPO increases (softer regions), the location of the minimum moves to higher energies, and the U shape becomes more pronounced (less elongated), with values of the phase lags at very low and very high energies becoming increasingly positive, close to 1 radian. 
The {\em U} shape of the phase lags shown in \hyperref[fig:rmslag]{Fig.~\ref{fig:rmslag}} is consistent with the plot in the middle panel of Fig.~11 in \cite{Konig-2024}.

We fitted a cubic polynomial to find the maximum or minimum of the phase-lag and rms spectra, and in each case we used MCMC simulations of $10^6$ samples to estimate the uncertainties of these values.
We report these values in \hyperref[tab:minmax]{Table~\ref{tab:minmax}}. It is now apparent that there is a positive relationship between the imaginary QPO frequency ($\nu_{\rm QPO}$) and the maximum of the rms spectrum, and the minimum of the phase-lag spectrum. As $\nu_{\rm QPO}$ increases from 1.5~Hz to 6~Hz, the energy at which the rms spectrum is maximum increases from 1.9 keV to 3.5 keV, while the energy at which the lag spectrum is minimum increases from 0.7~keV to 2~keV.


\subsection{On the structure of the ``plateau'' in the phase-lag frequency spectrum}
\label{sec:plateau}

As noted by \cite{Vaughan-1997}, the coherence function can be less than unity if more than one region of the accretion flow contributes to the signal in two energy bands, even if each region produces perfectly coherent variability. Indeed, if the observed light curve consists of two signals, each of them perfectly coherent in two energy bands, $i=1,2$, the coherence is less than unity except if $|Q_1| / |Q_2| = |R_1| / |R_2|$ and $\delta\theta_r = \delta\theta_q$, where $Q_i, R_i$ are the Fourier amplitudes of the two signals in the two bands, and $\delta\theta_r$ and $\delta\theta_q$ are the phase lags of the signals between those same two bands.

As seen in \hyperref[fig:pbubbles]{Fig.~\ref{fig:pbubbles}} and \hyperref[fig:plors]{Fig.~\ref{fig:plors}}, Lorentzians $L_6$ and $L_7$ of region Z3 (numbered by increasing characteristic frequency; see \hyperref[tab:plorstats]{Table~\ref{tab:plorstats}} for details)\footnote{In this work, the classical broad components used to fit the PDS of \cyg in the 0.1-10 Hz frequency range, $L_1$ and $L_2$  (see, e.g., \citealt{Pottschmidt-2003,Grinberg-2014,Konig-2024}), correspond to the combination of, respectively, Lorentzians $L_3+L_4$ and $L_6+L_7$.} have very similar ratios between the power spectra in the soft and hard energy bands (visual comparison), and very similar phase lags (${\sim}0.2$ rad). 
Specifically, we verified that the ratios $Q_1 / Q_2 \times R_2 / R_1$ and $\delta\theta_r / \delta\theta_q$ are all consistent with being 1 within 1$\sigma$. Here, $Q_{1,2}$ and $R_{1,2}$ represents the normalizations of respectively, $L_6$ and $L_7$ in the two energy bands, 0.3--2~keV and 2--12~keV band, respectively. 
All this leads to a plateau\footnote{In geological terms, a plateau is an area of flat terrain that is raised significantly above the surrounding area. Here we draw an analogy with the elevated phase lags that come after the cliff.} in the lag-frequency spectrum, at frequencies just above that of the cliff.

Given the best fit model of region Z3 shown in \hyperref[tab:zoneinfo]{Table~\ref{tab:zoneinfo}}, we linked the normalizations and lags of components $L_6$ and $L_7$, and we obtain a total $\chi^2$ of $1149.3$ for $1142$ degrees of freedom. An F-test of the two fits yields a null probability of $0.78$, which indicates that a model with more free parameters is not favoured.

We note then that the underlying structure behind this plateau can be approximated by a combination of two or more Lorentzian profiles with similar phase lags and correlated normalizations of the power spectra in the two energy bands by the relationships described above. Although in the previous Sections, and for the rest of this work, we let all parameters of each Lorentzian free, in this Section we have shown that some variability components may not be well described by a Lorentzian function, as the linear combination of Lorentzian functions is not a Lorentzian function.


\subsection{The high-frequency bump}
\label{sec:bump}

\begin{figure}
    \centering
    \includegraphics[width=0.99\linewidth]{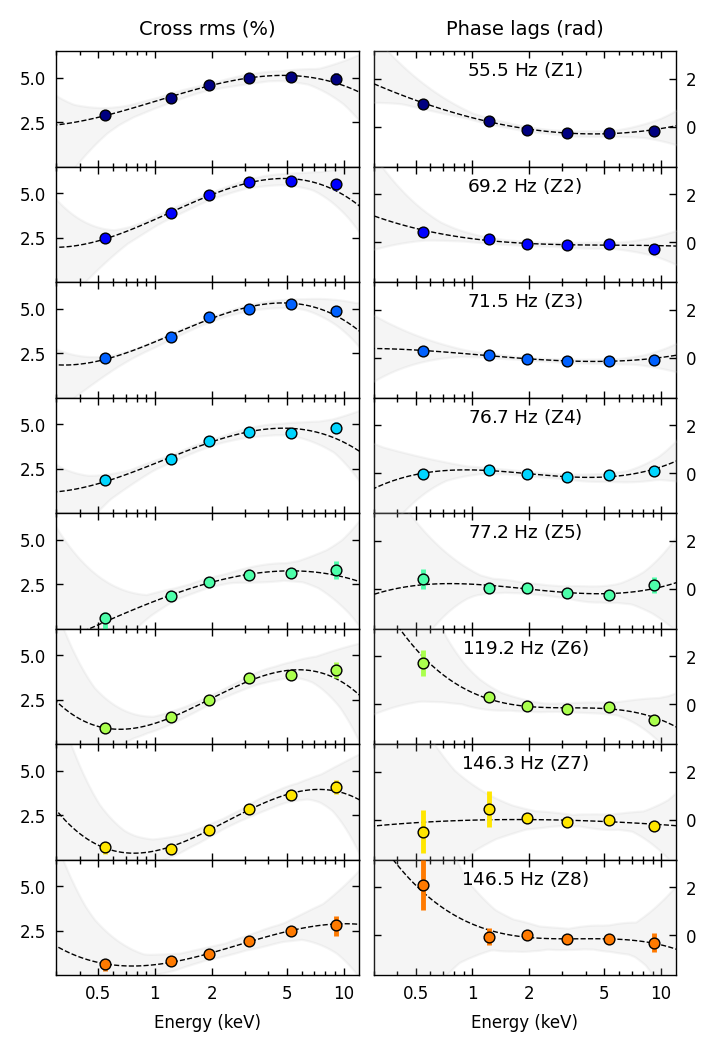}
    \caption{Same as \hyperref[fig:rmslag]{Fig.~\ref{fig:rmslag}} but for the high-frequency bump in \cyg.}
    \label{fig:bump}
\end{figure}

As seen in \hyperref[fig:plors]{Fig.~\ref{fig:plors}}, \cyg shows statistically significant variability at high frequency, up to ${\sim}200$~Hz; this variability, which can be fitted with a Lorentzian function, is present across a large part of the HID, with the centroid frequency of the best-fitting Lorentzian ranging from ${\sim}50$~Hz in the hardest region Z1 ($\Gamma{\simeq}1.4$) to ${\sim}160$~Hz in region Z8 ($\Gamma{\simeq}2.4$). This component is broad, with a quality factor between 0.2 and 0.5, and has a large negative phase lags of approximately $-0.5$ rad. As in previous work with sources like GRS~1915+105 \citep{Zhang-2022} and GX~339-4 \citep{Zhang2024}, we call this high-frequency variability component the HF bump.

We computed the covariance rms amplitude and phase-lag spectrum of the HF~bump across the 8 regions of the HID in which is significantly detected, which we show on \hyperref[fig:bump]{Fig.~\ref{fig:bump}}. 
In the same way as for the imaginary QPO, we fit the covariance rms and phase-lag spectrum in each region with a cubic polynomial (black line), and plot the $1\sigma$ uncertainty of the model as a grey shadow. Contrary to what we find for the imaginary QPO (\hyperref[fig:rmslag]{Fig.~\ref{fig:rmslag}}), neither the covariance rms nor the phase lags show any clear minimum/maximum.
The covariance rms amplitude of the HF bump does not exceed ${\sim}5$\%. The energy dependent phase lags tend towards zero at energies above 1~keV, and show small sign variations with no apparent pattern.

\begin{figure}
    \centering
    \includegraphics[width=0.99\linewidth]{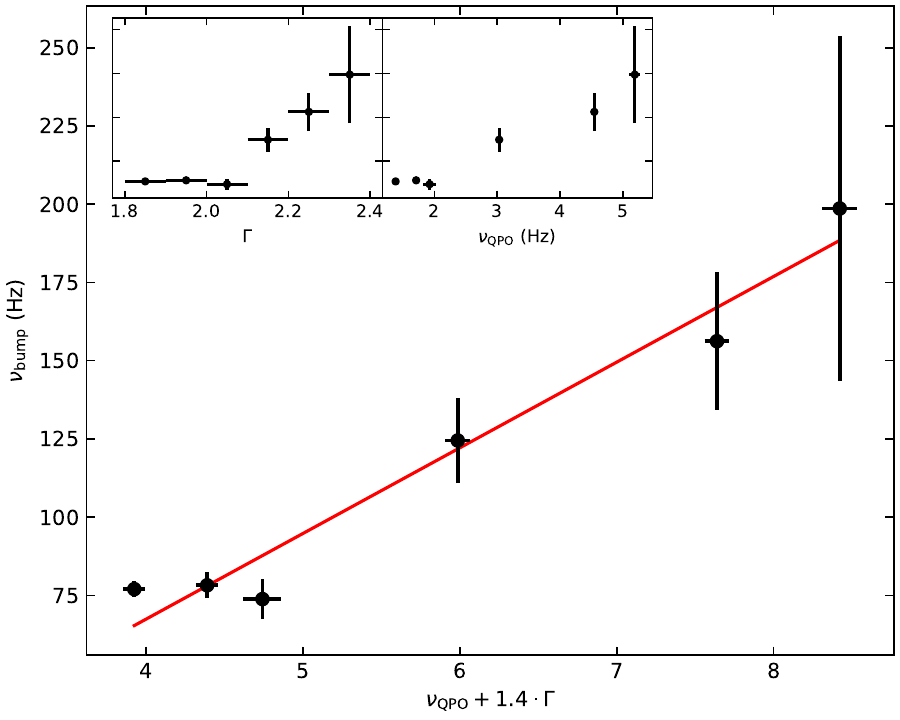}
    \caption{Characteristic frequency of the high-frequency bump of \cyg as a function of a linear combination of the spectral index, $\Gamma$, and the frequency of the imaginary QPO, $\nu_{\rm QPO}$. 
    Top insets show the individual correlations. The best fit linear model is shown in red.}
    \label{fig:bump_rotated}
\end{figure}

We note that the frequency of the HF bump ($\nu_{\rm bump}$) appears to follow a broken power-law relation with both the spectral index $\Gamma$ and the frequency of the imaginary QPO ($\nu_{\rm QPO}$). 
This is depicted in the insets of \hyperref[fig:bump_rotated]{Fig.~\ref{fig:bump_rotated}}. Following \cite{Zhang-2022}, we constructed a new variable $x=\nu_{\rm QPO}+c\cdot\Gamma$, where $c=\cot(\phi)$, with $\phi$ being the rotation angle of the $\nu_{\rm QPO}$ and $\Gamma$ axes around the axis of the frequency of the bump. 
The best ﬁt of $\log \nu_{\rm bump}$ vs. $x$ yields $c=1.4$, with a Pearson $R^2$ coefficient of $0.96$.
The relationship between the characteristic frequencies of the imaginary QPO and the HF bump with spectral hardness may indicate that these components have the same physical origin.


\section{Discussion}
\label{sec:discussion}

We report the discovery of a previously undetected quasi-periodic oscillation (QPO) in \cyg, which appears in intermediate spectral states of the source ($1.8 < \Gamma < 2.4$). This QPO is narrow ($Q \gtrsim 2$), with a large phase lag (${\sim}0.3$~rad), making it significant in the imaginary part of the cross-spectrum but not in the power spectrum. Because of this, we call it the imaginary QPO.
The centroid frequency of this QPO increases systematically from approximately ${\sim}1.5$ Hz to ${\sim}6$ Hz as \cyg transitions to softer spectral states. 
The rms energy spectrum of the imaginary QPO reveals a rising trend at higher energies, with a maximum that shifts from ${\sim}1.9$ keV when ${\Gamma}{\simeq}1.8$ to ${\sim}3.5$ keV when ${\Gamma}{\simeq}2.4$. Additionally, the phase-lag energy spectrum exhibits a characteristic {\em U} shape, with a local minimum that shifts from ${\sim}0.7$ keV when ${\Gamma}{\simeq}1.8$ ($kT_{\rm dbb}{\simeq}0.25$~keV) to ${\sim}2$ keV when ${\Gamma}{\simeq}2.4$ ($kT_{\rm dbb}{\simeq}0.35$~keV). 
The centroid frequency of the imaginary QPO also correlates closely with the centroid frequency of a high-frequency bump present in the hard and hard-intermediate spectral states of \cyg.

Recently, \cite{Konig-2024} reported a coherence drop and lag increase in the frequency range that lies in between the two well-known, broad  Lorentzian components (e.g., \citealt{Pottschmidt-2003}) in the power spectrum of \cyg, and other black hole binaries such as MAXI~J1820+070 and MAXI~J1348--630. 
\cite{Konig-2024} suggested that these features could be attributed both to the beating between two broad Lorentzian components, as well as an additional narrow Lorentzian component.
The need of a narrow Lorentzian component with large positive phase lags to fit simultaneously the PDS and CS supports the alternative interpretation that this imaginary QPO is an independent feature rather than a beat between the two broad Lorentzians. The beat between two broad components does not produce such a sharp drop in the coherence function as well as the steep rise in the phase lags 
This distinction is crucial for understanding the variability mechanisms in \cyg, as it indicates that multiple narrow components may dominate the broad-band variability structure in this system, even in the absence of significant features in the power spectrum \citep{Mendez-2023}.

We can compare our finding for \cyg with the results from the study of MAXI~J1820+070 (\citealt{Mendez-2023}; see also \citealt{Bellavita2025}) and MAXI~J1348--630 \citep{Alabarta2025}. 
\cite{Mendez-2023} and \cite{Alabarta2025} demonstrated that, while the PDS of, respectively, MAXI~J1820+070 and MAXI~J1348--630, could be modelled with four broad Lorentzian components, the joint fitting of the PDS and CS required at least seven narrower Lorentzian functions. This emphasises that the variability in the cross-spectrum can reveal hidden components that are not significantly detected in the power spectrum alone \citep{Mendez-2023}. In all these sources, the coherence drops when multiple Lorentzian components overlap in frequency, suggesting that the coherence function is sensitive to the detailed structure of the variability. In \cyg, we observed a similar effect, where the coherence drop and phase-lag cliff are linked to the presence of a narrow imaginary QPO component that is not significant in the PDS, but is significantly detected in the CS. 

The phase-lag and coherence spectra in \cyg resemble those observed in MAXI~J1348--630 during the decay of the 2019 outburst \citep{Alabarta2025}. 
Unlike in \cyg, where the imaginary QPO is only detected in the CS, the features in MAXI~J1348--630 were associated with a significant Lorentzian both in the PDS and CS. 
In MAXI~J1348--630, the sudden increase in the phase-lag spectrum (the cliff) and the narrow coherence drop happens at the same frequency as the type-C QPO, suggesting a direct link between the two features. 

Our findings on the imaginary QPO in \cyg align with known properties of type-C QPOs. The imaginary QPO appears only in hard-intermediate spectral states, spanning a frequency range from ${\sim}1$ Hz to ${\sim}6$ Hz, with a monotonic correlation between its frequency and the photon index of the Comptonizing component, $\Gamma$. The imaginary QPO also shows a rising rms energy trend, similar to type-C QPOs  (see e.g., \citealt{Alabarta-2022,Ma2023,Rawat-2023}), and exhibits phase-lag energy spectra with a distinctive ``U'' shape. Specifically, the phase-lag minimum correlates with the disc blackbody temperature (kT$_\mathrm{dbb}$; see \hyperref[tab:zoneinfo]{Table~\ref{tab:zoneinfo}} and \hyperref[tab:minmax]{Table~\ref{tab:minmax}}), consistent with predictions from the time-dependent Comptonization model vKompth \citep{Karpouzas-2021, Garcia-2021, Bellavita-2022}. These features, along with the results of \cite{Alabarta2025} in MAXI~J1348--630, strongly suggest that the imaginary QPO in \cyg is indeed a type-C QPO.

While a detailed application to Cygnus X-1 is beyond the scope of this paper, the observed phenomenology can still be interpreted within the framework of time-dependent Comptonization. In the vKompth model, soft photons from the accretion disc enter a spherical corona of hot electrons, undergoing inverse-Compton scattering and emerging with higher energies. Photons experiencing more scatterings escape later and at higher energies, producing hard lags. Conversely, if some up-scattered photons return to the disc, are reprocessed, and then re-emitted, this feedback mechanism introduces soft lags. The resulting phase-lag spectrum exhibits a characteristic ``U'' shape, with its minimum shifting to higher energies as the disc temperature increases (see Figure~2 of \citealt{Bellavita-2022}). Our results in \hyperref[tab:zoneinfo]{Table~\ref{tab:zoneinfo}} and \hyperref[tab:minmax]{Table~\ref{tab:minmax}} show that, as predicted by the model, the energy at which the ``U''-shaped phase-lag spectrum of the imaginary QPO is minimum increases as the temperature of the {\sc diskbb} component increases. This suggests that the imaginary QPO could be due to the time-dependent Comptonization.
 
An interesting comparison can be drawn between the imaginary QPO observed in \cyg and that seen in MAXI~J1820+070 and MAXI~J1348--630. In both cases, these sources exhibit QPOs that, in the PDS, are overshadowed by much stronger neighbouring variability components. This is in stark contrast to other black hole binaries, such as GRS~1915+105, GX~339--4 or Swift~J1727.8--1613, where confirmed type-C QPOs are stronger and stand out clearly in the power spectrum \citep{Zhang-2017, Belloni-2024, Mereminskiy-2024}. These stronger type-C QPOs, which are mainly observed in the high luminosity hard-to-soft transitions of an outburst, do not exhibit the coherence drop or the steep rise in phase lag that we observe in \cyg and MAXI~J1820+070. In contrast, the imaginary QPOs in MAXI~J1820+070 and MAXI~J1348--630 are observed in the soft-to-hard transition, during the decay of the outburst. This coincides with the region in the HID spanned by \cyg \citep{Konig-2024}. 
The presence of this imaginary QPO not only in a HMXB like \cyg, but also in other LMXBs, suggests that this variability component, with its associated phase-lag and coherence features, are not dependent upon the interaction of X-rays with the stellar wind of the companion, but are instead related to a phenomenon intrinsic to the accretion flow.
We emphasise that the soft state of \cyg, even on a scaled HID (see Fig.~1 of \citealt{Konig-2024}), does not coincide with the classical soft state of BH-LMXBs.

The fact that the same combination of Lorentzian functions that fits the PDS and CS can accurately predict the lag spectrum and the coherence function in \cyg (this paper) and other sources (MAXI J1820+070, \citealt{Mendez-2023,Bellavita2025}; MAXI~J1348--630, \citealt{Alabarta2025}) further supports the proposal that each Lorentzian component has its own distinct phase-lag frequency spectrum.
This idea challenges the view that broadband lags in X-ray binaries are due to a smooth global transfer function, such as in the case of reverberation \citep{Ingram-2009,IngramMastroserio2019} and propagating fluctuations \citep{Turner-2021,Mummery-2023}. Instead, the variability in these systems may arise from multiple, separate resonances within the accretion flow, each contributing to the observed lags and coherence behaviour. Our results provide further support to this picture, suggesting that the imaginary QPO we have detected could be one of these distinct resonances, reflecting a complex interplay between the accretion disc and the Comptonizing corona. Furthermore, not every Lorenztian component used to model the PDS and CS should be linked to a separate resonance. 
We showed in \hyperref[sec:plateau]{Sect.~\ref{sec:plateau}} that in many cases individual Lorenztians could represent different manifestations of a single variability component (e.g., fundamental plus harmonics), thus reducing the total number of separate resonances.

The model of \cite{Mendez-2023} assumes that each variability component is fully coherent across any two energy bands, while the different components themselves are mutually incoherent. Interestingly, the coherence drop in \cyg is only observed when including data below 2 keV, suggesting that the two variability components that cause the drop of the coherence may be the corona and the accretion disc. This interpretation seems counter-intuitive, as the disc provides seed photons for inverse Compton scattering in the corona, implying that variability in these two components should be correlated. One possible resolution to this apparent contradiction is if the disc variability responsible for the coherence drop arises from accretion rate fluctuations within the disc, which undergo viscous damping before reaching the corona \citep{Wilkinson2009}. Another possibility is that the low-energy variability is contributed by a component other than the disc.

Finally, we report, for the first time, a NICER detailed analysis of the high-frequency (HF) bump in \cyg. This HF component has been reported before using RXTE observations (see e.g., \citealt{Belloni1996, Nowak1999, Revnivtsev2000}), and is also apparent in the PDS in \cite{Konig-2024}, although they did not discuss it in detail.
This component, with frequencies ranging from 50 Hz to 150 Hz, exhibits large negative phase lags and shares similarities with other HF bumps observed in other black hole binaries. The discovery of this high-frequency component further reinforces the idea that multiple distinct Lorentzian components, each with its own timing and phase-lag properties, contribute to the overall variability of \cyg. This component and the imaginary QPO share similar properties with the high frequency bump and the type-C QPO detected in GRS~1915+105 \citep{Zhang-2022}. In both cases, the frequency of the HF bump is better correlated to a linear combination of the type-C QPO (the imaginary QPO in \cyg) and the spectral hardness ($\Gamma$ in the case of \cyg), than to each of those quantities separately. This reinforces the interpretation that the imaginary QPO in \cyg is the type-C QPO.



\begin{acknowledgements}

We are grateful to the anonymous reviewer for the comments that helped us improve the manuscript. This work is based on observations made by the NICER X-ray mission supported by NASA. This research has made use of data and software provided by the High Energy Astrophysics Science Archive Research Center (HEASARC), a service of the Astrophysics Science Division at NASA/GSFC and the High Energy Astrophysics Division of the Smithsonian Astrophysical Observatory. MM acknowledges the research programme Athena with project number 184.034.002, which is (partly) financed by the Dutch Research Council (NWO). FG is a CONICET researcher. FG acknowledges support by PIBAA 1275 and PIP 0113 (CONICET). FAF is a CONICET postdoc fellow. OK acknowledges NICER GO funding 80NSSC23K1660.

\end{acknowledgements}


\bibliographystyle{aa} 
\bibliography{biblio} 


\begin{appendix}


\section{Constant time-lag model}
\label{app:tlor}

As described in \cite{Mendez-2023}, the constant time-lag model (renamed $\tau$-model in this work) assumes, as the name suggests, that each Lorentzian component in the PDS and the CS has a frequency independent (constant) time lag. In this case, the argument of the trigonometric functions associated to every Lorentzian present in the real and imaginary parts of the cross-spectrum can be written as $2\pi\tau\nu$, where $\nu$ is the frequency and $\tau$ is the specific time lag. 

\begin{figure*}
    \centering
    \includegraphics[width=0.99\textwidth]{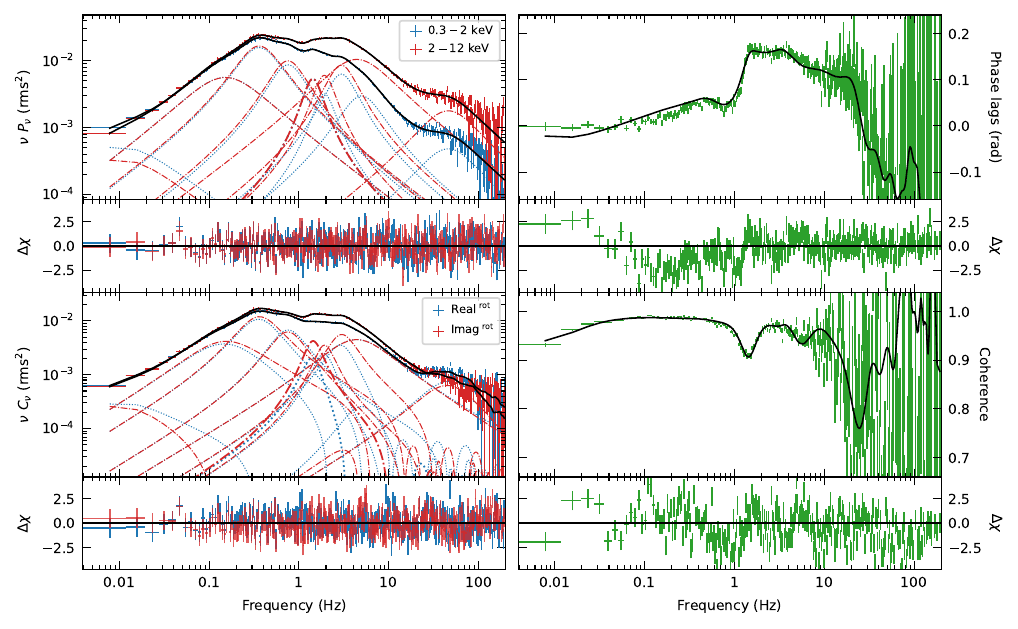}
    \caption{Best fit of the constant time-lag model applied to region Z3 of the HID of \cyg. Phase wrapping effects can be noticeable in the CS components, caused by the periodic zero-line crossing of the trigonometric functions. This effect is then propagated to the phase lags and the coherence function and becomes particularly noticeable at high frequencies.}
    \label{fig:tlor}
\end{figure*}

In \hyperref[fig:tlor]{Fig.~\ref{fig:tlor}} we present the best fit derived from the application of the $\tau$-model to region Z3. Phase wrapping, caused by the alternating sign change of the trigonometric functions for a given time lag as frequency increases, becomes apparent in the model of the real and imaginary part of the CS at frequencies above 20~Hz, although the quality of the data at those frequencies is not sufficient to reveal the presence of this effect. This effect is not present in the $\phi$-model (discussed in the main text) as the arguments of the trigonometric functions are independent of frequency.

\begin{figure}[h]
    \centering
    \includegraphics[width=0.99\columnwidth]{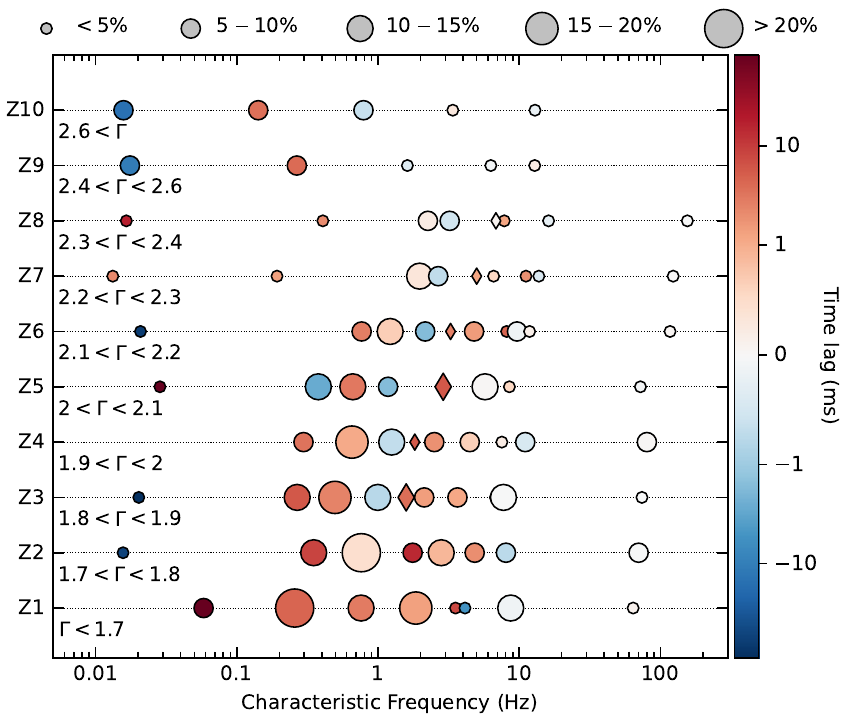}
    \caption{Covariance rms amplitude (\%) and time lags (ms) of \cyg derived from the $\tau$-model applied to the 10 regions of the HID. The diamond marker is used to trace the QPO that coincides with the coherence drop. See \hyperref[fig:pbubbles]{Fig.~\ref{fig:pbubbles}} for details.}
    \label{fig:tbub_tlag}
\end{figure}

In \hyperref[fig:tbub_tlag]{Fig.~\ref{fig:tbub_tlag}} we show the summary of the parameters of the Lorentzians in the $\tau$-model for the 10 regions in the HID in \hyperref[fig:hid]{Fig.~\ref{fig:hid}}.
Similarly to what is seen in \hyperref[fig:pbubbles]{Fig.~\ref{fig:pbubbles}}, the hard ($\Gamma<1.8$) and intermediate ($1.8<\Gamma<2.4$) regions, which correspond to regions Z1 to Z8, can be well described in the 0.004 to 200 Hz frequency range (almost 5 decades in frequency) by 8 Lorentzians that shift towards higher frequencies as the sources softens. The softest regions ($\Gamma>2.4$) can be described by 5 broad Lorentzians. 
At frequencies below 0.1~Hz, we obtain magnitudes of time lags greater than 10~ms. At frequencies between 0.1 and 10 Hz, the magnitude of the time lags range between 1 and 10 ms. At higher frequencies, the time-lags decrease towards zero.
Specifically, the QPO that coincides with the coherence drop has a positive time-lag close to 10~ms. The rms of this QPO is less than 10\%, reaching values below 5\% in some of the regions. In comparison, the same QPO in the $\phi$-model (see \hyperref[fig:pbubbles]{Fig.~\ref{fig:pbubbles}}) always had covariance rms amplitude lower than 5\%. We presume that the phase-wrapping effect can explain the disparity between the covariance rms amplitude measurements, as well as the total number of Lorentzian components that are statistically significant. 
Finally, in \hyperref[tab:stats]{Table~\ref{tab:stats}} we present the $\chi^2$ and the number of degrees of freedom associated to the fit of each region with the $\phi$- and $\tau$-models to the 10 regions of the HID, as well as the number of Lorentzian components used in each fit.

\label{app:taumodel}

\section{NICER dataset}
\label{app:nicerdata}

In \hyperref[tab:data]{Table~\ref{tab:data}} we show the full list of NICER observations of \cyg up to Cycle 6 used in this work. The dates correspond to the starting time of the observation. The exposure time column corresponds to the exposure of each observation after processing the data with {\tt nicerl2} and after computing the Fourier products with {\tt GHATS}, which discards time intervals that are shorter than the segment length of 128~s.

\begin{table*}[h!]
\renewcommand{\arraystretch}{1.15} 
\caption{\label{tab:data} NICER dataset properties up to Cycle 6 used in this work.}
\begin{center}
\begin{tabular}{c c l | c c l}
Obs.Id. & Start date (MJD) & Exposure (s) & Obs.Id. & Start date (MJD) & Exposure (s) \\ \hline
$0100320102$ & $57936.5$ & $1536$  & $5100320107$ & $59720.0$ & $2176$ \\  
$0100320103$ & $57937.1$ & $384$   & $5100320108$ & $59750.6$ & $2560$ \\
$0100320104$ & $57938.0$ & $9600$  & $5100320109$ & $59751.0$ & $9088$ \\
$0100320105$ & $57939.0$ & $12416$ & $5100320110$ & $59752.0$ & $8704$ \\ 
$0100320106$ & $57940.1$ & $7296$  & $5100320111$ & $59753.0$ & $3584$ \\ 
$0100320107$ & $57941.4$ & $3584$  & $5100320112$ & $59754.1$ & $2944$ \\ 
$0100320108$ & $57945.0$ & $384$   & $5100320113$ & $59755.1$ & $3200$ \\
$0100320109$ & $57945.0$ & $6016$  & $5100320114$ & $59756.0$ & $7808$ \\ 
$0100320110$ & $57946.1$ & $4480$  & $5100320115$ & $59757.0$ & $16128$ \\ 
$1100320101$ & $58054.3$ & $3456$  & $5100320116$ & $59758.0$ & $10880$ \\
$1100320102$ & $58055.0$ & $2944$  & $5100320117$ & $59759.0$ & $4864$ \\
$1100320103$ & $58056.0$ & $1280$  & $5100320118$ & $59760.0$ & $2048$ \\
$1100320106$ & $58088.9$ & $384$   & $5100320119$ & $59761.1$ & $768$ \\
$1100320107$ & $58089.1$ & $128$   & $5100320120$ & $59762.0$ & $1152$ \\
$1100320109$ & $58091.9$ & $128$   & $5100320121$ & $59763.4$ & $384$ \\
$1100320116$ & $58170.0$ & $384$   & $5100320122$ & $59764.6$ & $896$ \\ 
$1100320117$ & $58203.8$ & $6016$  & $5100320123$ & $59766.7$ & $1152$ \\ 
$1100320118$ & $58204.0$ & $4224$  & $5100320124$ & $59767.0$ & $1664$ \\
$1100320119$ & $58223.6$ & $11392$ & $5100320125$ & $59768.1$ & $640$ \\
$1100320120$ & $58224.0$ & $1792$  & $5100320126$ & $59772.1$ & $2944$ \\ 
$1100320121$ & $58265.3$ & $16512$ & $5100320127$ & $59773.3$ & $2816$ \\
$1100320122$ & $58341.1$ & $17664$ & $5100320128$ & $59774.3$ & $2176$ \\ 
$2636010101$ & $58701.4$ & $13312$ & $5100320129$ & $59775.1$ & $4224$ \\ 
$2636010102$ & $58800.4$ & $11648$ & $5100320130$ & $59776.0$ & $2560$ \\
$2636010201$ & $58745.0$ & $3840$  & $5100320131$ & $59777.0$ & $2688$ \\
$4690010104$ & $59642.4$ & $640$   & $5100320132$ & $59778.1$ & $1408$ \\
$4690010105$ & $59648.0$ & $1408$  & $5100320133$ & $59779.2$ & $896$ \\ 
$4690010106$ & $59653.5$ & $4224$  & $5100320134$ & $59906.1$ & $2816$ \\ 
$4690010107$ & $59658.4$ & $4736$  & $5100320139$ & $59913.9$ & $768$ \\
$4690010109$ & $59681.7$ & $3840$  & $5100320140$ & $59916.6$ & $2944$ \\
$4690010110$ & $59686.4$ & $3328$  & $5607010102$ & $59852.2$ & $384$ \\ 
$4690010111$ & $59687.0$ & $5120$  & $5607010103$ & $59854.5$ & $1152$ \\ 
$4690020101$ & $59622.8$ & $640$   & $5607010104$ & $59855.3$ & $1920$ \\ 
$4690020103$ & $59645.2$ & $640$   & $5607010202$ & $59858.0$ & $640$ \\
$4690020104$ & $59650.8$ & $896$   & $5607010203$ & $59860.1$ & $15744$ \\
$4690020105$ & $59656.2$ & $1792$  & $5607010204$ & $59861.0$ & $9600$ \\ 
$4690020109$ & $59678.9$ & $1920$  & $5607010301$ & $59863.5$ & $1408$ \\
$4690020110$ & $59684.4$ & $2432$  & $5607010302$ & $59865.6$ & $3712$ \\
$5100320101$ & $59714.3$ & $7680$  & $5607010303$ & $59866.1$ & $3328$ \\
$5100320102$ & $59715.1$ & $2432$  & $5607010401$ & $59869.7$ & $1664$ \\
$5100320103$ & $59716.0$ & $22272$ & $5607010402$ & $59871.6$ & $3968$ \\
$5100320104$ & $59717.0$ & $23680$ & $5607010403$ & $59872.0$ & $4864$ \\
$5100320105$ & $59718.0$ & $12928$ & $6643010101$ & $60088.7$ & $3840$ \\
$5100320106$ & $59719.0$ & $8832$  & $6643010102$ & $60089.0$ & $2048$ \\ 
 \hline
\end{tabular}
\end{center}
\tablefoot{The reported exposures refer to those remaining after processing the data with {\tt nicerl2} and computing the Fourier products with {\tt GHATS}, which discards time intervals that are shorter than the segment length of 128 s.}
\end{table*}

\section{Complete tables of Lorentzian properties}

In \hyperref[tab:plorstats]{Table~\ref{tab:plorstats}} and \hyperref[tab:tlorstats]{Table~\ref{tab:tlorstats}} we present the complete set of properties of the Lorentzian components used to fit the PDS and CS if the 10 regions of the HID of \cyg. See \hyperref[tab:stats]{Table~\ref{tab:stats}} for the statistical information on the best fit of each region and model. The $\star$ symbol indicates the row corresponding to the imaginary QPO, which is the core of this work.

\begin{table}[]
    \renewcommand{\arraystretch}{1.125}
    \centering
    \caption{\label{tab:plorstats} Phase-lag model summary.}
    \resizebox{0.99\columnwidth}{!}{
    \begin{tabular}{c c | r r r r r r }
        Region & Lor. & \multicolumn{1}{c}{$(\nu_0^{2}+\Delta^2)^{1/2}$} & \multicolumn{1}{c}{$Q\equiv\nu_0/\Delta$} & \multicolumn{1}{c}{Soft rms} & \multicolumn{1}{c}{Hard rms} & \multicolumn{1}{c}{CS rms} & \multicolumn{1}{c}{$\Delta\phi$} \\
               &            &   \multicolumn{1}{c}{Hz}           &   & \multicolumn{1}{c}{\%}       &   \multicolumn{1}{c}{\%}     &    \multicolumn{1}{c}{\%}  &   \multicolumn{1}{c}{rad}         \\
        \hline
        \parbox[t]{2mm}{\multirow{8}{*}{\rotatebox[origin=lb]{90}{\makecell[c]{Z1 \\ $\Gamma<1.7$}}}} & $1$ & $0.06{\pm}0.01$ & $0.04 {\pm}0.04 $ & $8 {\pm}0.8  $ & $4.9  {\pm}0.7  $ & $6.3  {\pm}0.8  $ & $-0.2 {\pm}0.07 $ \\
                            & $2$ & $0.26{\pm}0.01$ & $0.32 {\pm}0.04 $ & $22.6{\pm}0.8  $ & $18.8 {\pm}0.7  $ & $20.6 {\pm}0.8  $ & $0.05 {\pm}0.01 $ \\
                            & $3$ & $0.8 {\pm}0.04$ & $0.44 {\pm}0.07 $ & $16  {\pm}1  $ & $15 {\pm}1  $ & $15 {\pm}1  $ & $0.1  {\pm}0.01 $ \\
                            & $4$ & $1.54{\pm}0.05$ & $0.84 {\pm}0.08 $ & $11  {\pm}1  $ & $12 {\pm}1  $ & $12 {\pm}1  $ & $0.12 {\pm}0.02 $ \\
                            & $5$ & $3.5 {\pm}0.2 $ & $0.39 {\pm}0.06 $ & $12.2{\pm}0.9  $ & $13 {\pm}1  $ & $12 {\pm}1  $ & $0.3  {\pm}0.04 $ \\
                            & $6$ & $6.9 {\pm}0.4 $ & $0.5  {\pm}0.06 $ & $5.5 {\pm}0.8  $ & $8  {\pm}1  $ & $7.4  {\pm}0.7  $ & $-0.3 {\pm}0.1  $ \\
                            & $7$ & $10.9{\pm}0.4 $ & $0.49 {\pm}0.06 $ & $2.6 {\pm}0.7  $ & $8.1  {\pm}0.8  $ & $5.6  {\pm}0.5  $ & $0.6  {\pm}0.1  $ \\
                            & $8$ & $56  {\pm}2$    & $0.31 {\pm}0.06 $ & $3.5 {\pm}0.2  $ & $5.8  {\pm}0.3  $ & $4.3  {\pm}0.2  $ & $-0.47{\pm}0.05 $ \\
        \hline
        \parbox[t]{2mm}{\multirow{8}{*}{\rotatebox[origin=lb]{90}{\makecell[c]{Z2 \\ $1.7<\Gamma<1.8$}}}} & $1$ & $0.014{\pm}0.005$ & $0.1  {\pm}0.2  $ & $3.8  {\pm}0.4  $ & $1.7  {\pm}0.4  $ & $2.7  {\pm}0.3  $ & $-0.4 {\pm}0.1  $ \\
                            & $2$ & $0.5  {\pm}0.02 $ & $0.14 {\pm}0.02 $ & $23.8 {\pm}0.5  $ & $20.8 {\pm}0.5  $ & $22.3 {\pm}0.5  $ & $0.031{\pm}0.008$ \\
                            & $3$ & $1.5  {\pm}0.1  $ & $0.3  {\pm}0.06 $ & $12 {\pm}1  $     & $13 {\pm}1  $ & $13 {\pm}1  $ & $0.14 {\pm}0.03 $ \\
                            & $4$ & $2.6  {\pm}0.1  $ & $0.62 {\pm}0.06 $ & $10.9 {\pm}0.9  $ & $14 {\pm}1  $ & $12 {\pm}1  $ & $0.21 {\pm}0.02 $ \\
                            & $5$ & $4.5  {\pm}0.2  $ & $0.8  {\pm}0.09 $ & $6.6  {\pm}0.8  $ & $9  {\pm}1  $ & $7.6  {\pm}0.9  $ & $0.21 {\pm}0.03 $ \\
                            & $6$ & $7.4  {\pm}0.3  $ & $1  {\pm}0.2  $ & $2.5  {\pm}0.5  $ & $5.9  {\pm}0.7  $ & $4.4  {\pm}0.6  $ & $-0.06{\pm}0.06 $ \\
                            & $7$ & $13.9 {\pm}0.8  $ & $0.7  {\pm}0.1  $ & $0.5  {\pm}0.3  $ & $5.6  {\pm}0.7  $ & $3.2  {\pm}0.4  $ & $0.4  {\pm}0.1  $ \\
                            & $8$ & $72   {\pm}4  $   & $0.22 {\pm}0.07 $ & $3.3  {\pm}0.2  $ & $6.9  {\pm}0.3  $ & $4.6  {\pm}0.2  $ & $-0.2 {\pm}0.04 $ \\
        \hline
        \parbox[t]{2mm}{\multirow{10}{*}{\rotatebox[origin=lb]{90}{\makecell[c]{Z3 \\ $1.8<\Gamma<1.9$}}}} & $1$  & $0.017{\pm}0.003$ & $0.1  {\pm}0.1  $ & $3.7  {\pm}0.3  $ & $2.9  {\pm}0.3  $ & $3.2  {\pm}0.2  $ & $-0.01{\pm}0.07 $ \\
                             & $2$  & $0.32 {\pm}0.03 $ & $0.22 {\pm}0.03 $ & $12.8 {\pm}0.7  $ & $12.8 {\pm}0.7  $ & $12.8 {\pm}0.7  $ & $-0.03{\pm}0.01 $ \\
                             & $3$  & $0.5  {\pm}0.02 $ & $0.76 {\pm}0.06 $ & $14.8 {\pm}0.8  $ & $15.2 {\pm}0.9  $ & $15   {\pm}0.8  $ & $0.08 {\pm}0.01 $ \\
                             & $4$  & $1.03 {\pm}0.02 $ & $0.81 {\pm}0.07 $ & $11.4 {\pm}0.7  $ & $12.7 {\pm}0.7  $ & $12   {\pm}0.7  $ & $-0.02{\pm}0.01 $ \\
                             & $5^\star$  & $1.51 {\pm}0.04 $ & $2.3  {\pm}0.3  $ & $5    {\pm}0.6  $ & $5.1  {\pm}0.8  $ & $4.7  {\pm}0.7  $ & $0.37 {\pm}0.04 $ \\
                             & $6$  & $2.17 {\pm}0.06 $ & $1.4  {\pm}0.2  $ & $6.4  {\pm}0.6  $ & $9.1  {\pm}0.8  $ & $7.7  {\pm}0.7  $ & $0.21 {\pm}0.02 $ \\
                             & $7$  & $3.64 {\pm}0.07 $ & $1.17 {\pm}0.08 $ & $8.1  {\pm}0.5  $ & $11.5 {\pm}0.7  $ & $9.6  {\pm}0.6  $ & $0.194{\pm}0.009$ \\
                             & $8$  & $6.6  {\pm}0.1  $ & $0.83 {\pm}0.03 $ & $6.5  {\pm}0.4  $ & $10   {\pm}0.6  $ & $8    {\pm}0.5  $ & $0.14 {\pm}0.01 $ \\
                             & $9$  & $13.2 {\pm}0.5  $ & $0.61 {\pm}0.08 $ & $0.7  {\pm}0.4  $ & $6.9  {\pm}0.4  $ & $3.8  {\pm}0.3  $ & $0.09 {\pm}0.04 $ \\
                             & $10$ & $77   {\pm}2  $   & $0.46 {\pm}0.04 $ & $3.1  {\pm}0.07 $ & $6.6  {\pm}0.2  $ & $4.3  {\pm}0.1  $ & $-0.21{\pm}0.03 $ \\
        \hline
        \parbox[t]{2mm}{\multirow{9}{*}{\rotatebox[origin=lb]{90}{\makecell[c]{Z4 \\ $1.9<\Gamma<2$}}}} & $1$ & $0.31 {\pm}0.04 $ & $0.02 {\pm}0.02 $ & $10 {\pm}0.9  $ & $11 {\pm}1  $ & $10.4 {\pm}0.9  $ & $-0.06{\pm}0.03 $ \\
                             & $2$ & $0.65 {\pm}0.02 $ & $0.53 {\pm}0.04 $ & $17.4 {\pm}0.7  $ & $19.6 {\pm}0.9  $ & $18.5 {\pm}0.8  $ & $0.06 {\pm}0.01 $ \\
                             & $3$ & $1.27 {\pm}0.04 $ & $0.84 {\pm}0.09 $ & $9.1  {\pm}0.8  $ & $11.6 {\pm}0.8  $ & $10.2 {\pm}0.8  $ & $-0.05{\pm}0.02 $ \\
                             & $4^\star$ & $1.81 {\pm}0.04 $ & $2.9  {\pm}0.4  $ & $3.9  {\pm}0.5  $ & $3.8  {\pm}0.7  $ & $3.6  {\pm}0.5  $ & $0.54 {\pm}0.08 $ \\
                             & $5$ & $2.49 {\pm}0.06 $ & $1.7  {\pm}0.2  $ & $5.2  {\pm}0.5  $ & $8.1  {\pm}0.8  $ & $6.6  {\pm}0.6  $ & $0.29 {\pm}0.03 $ \\
                             & $6$ & $4.5  {\pm}0.1  $ & $1.08 {\pm}0.08 $ & $7.7  {\pm}0.4  $ & $12.5 {\pm}0.7  $ & $9.8  {\pm}0.6  $ & $0.19 {\pm}0.01 $ \\
                             & $7$ & $7.6  {\pm}0.1  $ & $1.1  {\pm}0.1  $ & $4.1  {\pm}0.4  $ & $4  {\pm}1  $ & $4.4  {\pm}0.6  $ & $0.21 {\pm}0.06 $ \\
                             & $8$ & $11.1 {\pm}0.9  $ & $0.51 {\pm}0.06 $ & $0.7  {\pm}0.4  $ & $9.3  {\pm}0.9  $ & $5  {\pm}0.5  $ & $0.02 {\pm}0.05 $ \\
                             & $9$ & $78 {\pm}4  $ & $0.5  {\pm}0.06 $ & $2.77 {\pm}0.09 $ & $6.1  {\pm}0.2  $ & $3.9  {\pm}0.1  $ & $-0.21{\pm}0.04 $ \\
        \hline
        \parbox[t]{2mm}{\multirow{9}{*}{\rotatebox[origin=lb]{90}{\makecell[c]{Z5 \\ $2<\Gamma<2.1$}}}}  & $1$ & $0.55 {\pm}0.04 $ & $0.01 {\pm}0.01 $ & $12.1 {\pm}0.5  $ & $15 {\pm}0.7  $ & $13.5 {\pm}0.6  $ & $0 {\pm}0.02 $ \\
                             & $2$ & $0.59 {\pm}0.02 $ & $1.2  {\pm}0.1  $ & $9.4  {\pm}0.8  $ & $12 {\pm}0.9  $ & $10.6 {\pm}0.8  $ & $0.08 {\pm}0.03 $ \\
                             & $3$ & $1.22 {\pm}0.03 $ & $0.9  {\pm}0.1  $ & $9.6  {\pm}0.7  $ & $12 {\pm}1  $ & $10.7 {\pm}0.8  $ & $0.02 {\pm}0.03 $ \\
                             & $4^\star$ & $2.2  {\pm}0.1  $ & $1.7  {\pm}0.3  $ & $5.1  {\pm}0.6  $ & $2  {\pm}1  $ & $4.2  {\pm}0.7  $ & $0.4  {\pm}0.1  $ \\
                             & $5$ & $3.2  {\pm}0.1  $ & $1.4  {\pm}0.2  $ & $4.9  {\pm}0.7  $ & $5.7  {\pm}0.9  $ & $5.7  {\pm}0.7  $ & $0.51 {\pm}0.06 $ \\
                             & $6$ & $5.2  {\pm}0.2  $ & $1.4  {\pm}0.2  $ & $5.2  {\pm}0.6  $ & $7  {\pm}1  $ & $6.3  {\pm}0.8  $ & $0.31 {\pm}0.04 $ \\
                             & $7$ & $5.9  {\pm}0.4  $ & $0.22 {\pm}0.07 $ & $1.4  {\pm}0.8  $ & $15.9 {\pm}0.7  $ & $9  {\pm}0.6  $ & $-0.18{\pm}0.07 $ \\
                             & $8$ & $8.5  {\pm}0.3  $ & $1.1  {\pm}0.1  $ & $3.6  {\pm}0.6  $ & $4.8  {\pm}0.9  $ & $4.1  {\pm}0.7  $ & $0.34 {\pm}0.07 $ \\
                             & $9$ & $73 {\pm}6  $ & $0.8  {\pm}0.2  $ & $1.6  {\pm}0.2  $ & $4.4  {\pm}0.3  $ & $2.5  {\pm}0.2  $ & $-0.1 {\pm}0.1  $ \\
        \hline
        \parbox[t]{2mm}{\multirow{8}{*}{\rotatebox[origin=lb]{90}{\makecell[c]{Z6 \\ $2.1<\Gamma<2.2$}}}}  & $1$ & $0.022{\pm}0.005$ & $0.04 {\pm}0.04 $ & $2.2  {\pm}0.1  $ & $3.4  {\pm}0.2  $ & $2.7  {\pm}0.1  $ & $-0.01{\pm}0.08 $ \\
                             & $2$ & $0.71 {\pm}0.09 $ & $0.09 {\pm}0.05 $ & $7  {\pm}0.6  $ & $10.7 {\pm}0.9  $ & $8.6  {\pm}0.7  $ & $-0.05{\pm}0.03 $ \\
                             & $3$ & $1.17 {\pm}0.04 $ & $0.64 {\pm}0.05 $ & $10.3 {\pm}0.5  $ & $15.9 {\pm}0.8  $ & $12.7 {\pm}0.7  $ & $0.07 {\pm}0.02 $ \\
                             & $4$ & $2.19 {\pm}0.06 $ & $0.9  {\pm}0.1  $ & $6  {\pm}0.5  $ & $10.5 {\pm}0.8  $ & $7.9  {\pm}0.7  $ & $-0.11{\pm}0.03 $ \\
                             & $5^\star$ & $3.3  {\pm}0.1  $ & $2.2  {\pm}0.4  $ & $3.2  {\pm}0.4  $ & $3.5  {\pm}0.8  $ & $3.3  {\pm}0.5  $ & $0.29 {\pm}0.07 $ \\
                             & $6$ & $5.3  {\pm}0.2  $ & $1.1  {\pm}0.1  $ & $4.7  {\pm}0.3  $ & $10.9 {\pm}0.7  $ & $7.3  {\pm}0.4  $ & $0.31 {\pm}0.02 $ \\
                             & $7$ & $10.2 {\pm}0.2  $ & $0.85 {\pm}0.05 $ & $3.6  {\pm}0.3  $ & $9.1  {\pm}0.7  $ & $5.6  {\pm}0.4  $ & $0.03 {\pm}0.03 $ \\
                             & $8$ & $120{\pm}10 $ & $0.1  {\pm}0.1  $ & $1.1  {\pm}0.2  $ & $5.6  {\pm}0.3  $ & $2.8  {\pm}0.2  $ & $-0.33{\pm}0.08 $ \\
        \hline
        \parbox[t]{2mm}{\multirow{9}{*}{\rotatebox[origin=lb]{90}{\makecell[c]{Z7 \\ $2.2<\Gamma<2.3$}}}}  & $1$ & $0.014{\pm}0.002$ & $0.06 {\pm}0.08 $ & $2.9  {\pm}0.2  $ & $4.1  {\pm}0.3  $ & $3.3  {\pm}0.2  $ & $0.03 {\pm}0.07 $ \\
                             & $2$ & $0.21 {\pm}0.03 $ & $0.02 {\pm}0.03 $ & $2.8  {\pm}0.1  $ & $4.9  {\pm}0.2  $ & $3.7  {\pm}0.2  $ & $-0.06{\pm}0.05 $ \\
                             & $3$ & $1.96 {\pm}0.07 $ & $0.35 {\pm}0.02 $ & $9.8  {\pm}0.3  $ & $17.3 {\pm}0.7  $ & $13 {\pm}0.4  $ & $0.03 {\pm}0.01 $ \\
                             & $4$ & $2.74 {\pm}0.06 $ & $0.78 {\pm}0.08 $ & $4.8  {\pm}0.6  $ & $10 {\pm}1  $ & $7  {\pm}0.8  $ & $-0.11{\pm}0.03 $ \\
                             & $5^\star$ & $5.1  {\pm}0.1  $ & $1.9  {\pm}0.1  $ & $3.3  {\pm}0.2  $ & $6  {\pm}0.7  $ & $4.3  {\pm}0.4  $ & $0.23 {\pm}0.03 $ \\
                             & $6$ & $7.3  {\pm}0.4  $ & $1.2  {\pm}0.1  $ & $1.5  {\pm}0.4  $ & $7.5  {\pm}0.7  $ & $4.1  {\pm}0.4  $ & $0.27 {\pm}0.04 $ \\
                             & $7$ & $11 {\pm}0.3  $ & $1.6  {\pm}0.2  $ & $2  {\pm}0.1  $ & $4.3  {\pm}0.6  $ & $2.8  {\pm}0.3  $ & $0.1  {\pm}0.05 $ \\
                             & $8$ & $15.5 {\pm}0.5  $ & $1.7  {\pm}0.2  $ & $0.5  {\pm}0.2  $ & $3.3  {\pm}0.3  $ & $1.7  {\pm}0.2  $ & $-0.3 {\pm}0.1  $ \\
                             & $9$ & $150{\pm}20 $ & $0.2  {\pm}0.1  $ & $0.4  {\pm}0.2  $ & $5  {\pm}0.3  $ & $2  {\pm}0.2  $ & $0 {\pm}0.1  $ \\
        \hline
        \parbox[t]{2mm}{\multirow{8}{*}{\rotatebox[origin=lb]{90}{\makecell[c]{Z8 \\ $2.3<\Gamma<2.4$}}}}  & $1$ & $0.017{\pm}0.002$ & $0.04 {\pm}0.05 $ & $1.93 {\pm}0.09 $ & $3.7  {\pm}0.2  $ & $2.5  {\pm}0.1  $ & $-0.04{\pm}0.06 $ \\
                             & $2$ & $0.42 {\pm}0.06 $ & $0.02 {\pm}0.02 $ & $2.3  {\pm}0.1  $ & $5  {\pm}0.3  $ & $3.3  {\pm}0.2  $ & $0.03 {\pm}0.03 $ \\
                             & $3$ & $2.2  {\pm}0.1  $ & $0.45 {\pm}0.03 $ & $6.3  {\pm}0.3  $ & $13.1 {\pm}0.6  $ & $9.1  {\pm}0.4  $ & $0.02 {\pm}0.01 $ \\
                             & $4$ & $3.34 {\pm}0.07 $ & $0.94 {\pm}0.08 $ & $3.9  {\pm}0.4  $ & $9.5  {\pm}0.7  $ & $6.1  {\pm}0.5  $ & $-0.07{\pm}0.02 $ \\
                             & $5^\star$ & $6.1  {\pm}0.2  $ & $1.6  {\pm}0.2  $ & $2.4  {\pm}0.3  $ & $5  {\pm}0.9  $ & $3.5  {\pm}0.5  $ & $0.01 {\pm}0.06 $ \\
                             & $6$ & $8.4  {\pm}0.4  $ & $1.1  {\pm}0.1  $ & $2  {\pm}0.3  $ & $6.9  {\pm}0.8  $ & $3.8  {\pm}0.5  $ & $0.34 {\pm}0.06 $ \\
                             & $7$ & $15.8 {\pm}0.3  $ & $1.3  {\pm}0.1  $ & $1.2  {\pm}0.1  $ & $4.2  {\pm}0.4  $ & $2.3  {\pm}0.2  $ & $-0.1 {\pm}0.05 $ \\
                             & $8$ & $190{\pm}50 $ & $0.3  {\pm}0.2  $ & $0.5  {\pm}0.2  $ & $3.6  {\pm}0.4  $ & $1.4  {\pm}0.2  $ & $-0.4 {\pm}0.3  $ \\
        \hline
        \parbox[t]{2mm}{\multirow{6}{*}{\rotatebox[origin=lb]{90}{\makecell[c]{Z9 \\ $2.4<\Gamma<2.6$}}}}  & $1$ & $0.018{\pm}0.001$ & $0.05 {\pm}0.05 $ & $5  {\pm}0.2  $ & $7.9  {\pm}0.2  $ & $5.6  {\pm}0.2  $ & $0.01 {\pm}0.04 $ \\
                             & $2$ & $0.27 {\pm}0.01 $ & $0.03 {\pm}0.03 $ & $4.4  {\pm}0.1  $ & $7.2  {\pm}0.2  $ & $5.2  {\pm}0.1  $ & $0.01 {\pm}0.02 $ \\
                             & $3$ & $1.6  {\pm}0.1  $ & $0.04 {\pm}0.04 $ & $3.1  {\pm}0.1  $ & $6.9  {\pm}0.2  $ & $4.5  {\pm}0.2  $ & $0.05 {\pm}0.02 $ \\
                             & $4$ & $6.2  {\pm}0.2  $ & $0.4  {\pm}0.03 $ & $3  {\pm}0.1  $ & $8.1  {\pm}0.4  $ & $5  {\pm}0.2  $ & $0  {\pm}0.02 $ \\
                             & $5$ & $9.7  {\pm}0.8  $ & $0.5  {\pm}0.1  $ & $0.6  {\pm}0.3  $ & $4.8  {\pm}0.6  $ & $1.9  {\pm}0.3  $ & $0.03 {\pm}0.09 $ \\
                             & $6$ & $25 {\pm}2  $ & $0.8  {\pm}0.2  $ & $0.4  {\pm}0.1  $ & $2.7  {\pm}0.4  $ & $1.1  {\pm}0.2  $ & $-0.3 {\pm}0.1  $ \\
        \hline
        \parbox[t]{2mm}{\multirow{6}{*}{\rotatebox[origin=lb]{90}{\makecell[c]{Z10 \\ $\Gamma>2.6$}}}} & $1$ & $0.017{\pm}0.002$ & $0.02 {\pm}0.02 $ & $6.6  {\pm}0.2  $ & $9  {\pm}0.2  $ & $6.7  {\pm}0.2  $ & $-0.03{\pm}0.04 $ \\
                             & $2$ & $0.16 {\pm}0.01 $ & $0.01 {\pm}0.01 $ & $6  {\pm}0.1  $ & $8.6  {\pm}0.2  $ & $6.6  {\pm}0.2  $ & $0.01 {\pm}0.02 $ \\
                             & $3$ & $0.91 {\pm}0.05 $ & $0.05 {\pm}0.04 $ & $4.6  {\pm}0.2  $ & $8.2  {\pm}0.3  $ & $5.8  {\pm}0.2  $ & $-0.02{\pm}0.01 $ \\
                             & $4$ & $3.7  {\pm}0.2  $ & $0.19 {\pm}0.05 $ & $1.9  {\pm}0.2  $ & $6.1  {\pm}0.3  $ & $3.5  {\pm}0.2  $ & $0.06 {\pm}0.02 $ \\
                             & $5$ & $9.6  {\pm}0.5  $ & $0.56 {\pm}0.08 $ & $0.6  {\pm}0.1  $ & $3.5  {\pm}0.3  $ & $1.7  {\pm}0.2  $ & $0.06 {\pm}0.04 $ \\
                             & $6$ & $20 {\pm}1  $ & $1  {\pm}0.2  $ & $0.2  {\pm}0.08 $ & $1.9  {\pm}0.2  $ & $0.73 {\pm}0.09 $ & $-0.9 {\pm}0.2  $ \\
        \hline
    \end{tabular}
    }
    \tablefoot{Properties of the Lorentzian components of the $\phi$-model for the 10 regions of the HID of \cyg. The $\star$ symbol indicates the row corresponding to the imaginary QPO. Soft (Hard) rms corresponds to the root square of the normalization of the Lorentzian in the 0.3--2~keV (2--12~~keV) energy band. CS rms is the fractional covariance rms amplitude of the Lorentzian, and $\Delta\phi$ is the phase-lag of the Lorentzian.}
\end{table}

\begin{table}[]
    \renewcommand{\arraystretch}{1.175}
    \centering
    \caption{\label{tab:tlorstats} Time-lag model summary.}
    \resizebox{0.99\columnwidth}{!}{
    \begin{tabular}{c c | r r r r r r }
        Region & Lor. & \multicolumn{1}{c}{$(\nu_0^{2}+\Delta^2)^{1/2}$} & \multicolumn{1}{c}{$Q\equiv\nu_0/\Delta$} & \multicolumn{1}{c}{Soft rms} & \multicolumn{1}{c}{Hard rms} & \multicolumn{1}{c}{CS rms} & \multicolumn{1}{c}{$\Delta t$} \\
               &            &   \multicolumn{1}{c}{Hz}           &   & \multicolumn{1}{c}{\%}       &   \multicolumn{1}{c}{\%}     &    \multicolumn{1}{c}{\%}  &   \multicolumn{1}{c}{ms}         \\
        \hline
\parbox[t]{2mm}{\multirow{8}{*}{\rotatebox[origin=lb]{90}{\makecell[c]{Z1 \\ $\Gamma<1.7$}}}} & $1$ & $0.058{\pm}0.007$ & $0.03 {\pm}0.03 $ & $8.2  {\pm}0.5  $ & $5  {\pm}0.5  $ & $6.3  {\pm}0.5  $ & $830{\pm}30 $ \\
                     & $2$ & $0.257{\pm}0.008$ & $0.32 {\pm}0.02 $ & $22.4 {\pm}0.5  $ & $18.7 {\pm}0.4  $ & $20.5 {\pm}0.4  $ & $44 {\pm}1  $ \\
                     & $3$ & $0.76 {\pm}0.03 $ & $0.45 {\pm}0.04 $ & $15.7 {\pm}0.7  $ & $13.6 {\pm}0.6  $ & $14.8 {\pm}0.7  $ & $29 {\pm}1  $ \\
                     & $4$ & $1.85 {\pm}0.03 $ & $0.56 {\pm}0.03 $ & $15.4 {\pm}0.4  $ & $16.2 {\pm}0.5  $ & $15.5 {\pm}0.4  $ & $13.3 {\pm}0.5  $ \\
                     & $5$ & $3.5  {\pm}0.3  $ & $0.7  {\pm}0.1  $ & $4  {\pm}1  $ & $4.3  {\pm}0.7  $ & $3.3  {\pm}0.3  $ & $78 {\pm}4  $ \\
                     & $6$ & $4.1  {\pm}0.1  $ & $0.78 {\pm}0.06 $ & $5.5  {\pm}0.9  $ & $1.4  {\pm}0.8  $ & $4.6  {\pm}0.2  $ & $-53{\pm}2  $ \\
                     & $7$ & $8.7  {\pm}0.2  $ & $0.25 {\pm}0.03 $ & $7.6  {\pm}0.3  $ & $14 {\pm}0.4  $ & $13.3 {\pm}0.3  $ & $-1.2 {\pm}0.2  $ \\
                     & $8$ & $64 {\pm}4  $ & $0.35 {\pm}0.09 $ & $3.1  {\pm}0.2  $ & $4.9  {\pm}0.3  $ & $4.2  {\pm}0.3  $ & $1  {\pm}0.2  $ \\
\hline
\parbox[t]{2mm}{\multirow{8}{*}{\rotatebox[origin=lb]{90}{\makecell[c]{Z2 \\ $1.7<\Gamma<1.8$}}}}  & $1$ & $0.016{\pm}0.005$ & $0.1  {\pm}0.1  $ & $3.9  {\pm}0.3  $ & $1.8  {\pm}0.4  $ & $2.7  {\pm}0.3  $ & $(-1{\pm}2){\cdot}10^3$ \\
                     & $2$ & $0.35 {\pm}0.01 $ & $0.29 {\pm}0.03 $ & $14.3 {\pm}0.7  $ & $11.5 {\pm}0.8  $ & $12.6 {\pm}0.6  $ & $88 {\pm}2  $ \\
                     & $3$ & $0.76 {\pm}0.04 $ & $0.02 {\pm}0.02 $ & $21.7 {\pm}0.6  $ & $20.1 {\pm}0.6  $ & $21 {\pm}0.4  $ & $5  {\pm}1  $ \\
                     & $4$ & $1.8  {\pm}0.2  $ & $0.42 {\pm}0.07 $ & $6  {\pm}1  $ & $6  {\pm}1  $ & $5.1  {\pm}0.3  $ & $146{\pm}3  $ \\
                     & $5$ & $2.8  {\pm}0.07 $ & $0.5  {\pm}0.03 $ & $12 {\pm}0.5  $ & $14.5 {\pm}0.7  $ & $14 {\pm}0.5  $ & $9  {\pm}0.7  $ \\
                     & $6$ & $4.8  {\pm}0.1  $ & $0.76 {\pm}0.05 $ & $5.9  {\pm}0.5  $ & $1.7  {\pm}0.8  $ & $5.6  {\pm}0.4  $ & $19 {\pm}1  $ \\
                     & $7$ & $8.1  {\pm}0.4  $ & $0.2  {\pm}0.03 $ & $1  {\pm}0.5  $ & $13.1 {\pm}0.6  $ & $8.8  {\pm}0.5  $ & $-7.5 {\pm}0.4  $ \\
                     & $8$ & $70 {\pm}4  $ & $0.34 {\pm}0.05 $ & $2.8  {\pm}0.1  $ & $5.5  {\pm}0.3  $ & $6.4  {\pm}0.2  $ & $-0.2 {\pm}0.1  $ \\
\hline
\parbox[t]{2mm}{\multirow{10}{*}{\rotatebox[origin=lb]{90}{\makecell[c]{Z3 \\ $1.8<\Gamma<1.9$}}}} & $1$ & $0.02 {\pm}0.003$ & $0.06 {\pm}0.07 $ & $4  {\pm}0.2  $ & $3.2  {\pm}0.2  $ & $3.5  {\pm}0.2  $ & $-900{\pm}400$ \\
                     & $2$ & $0.27 {\pm}0.01 $ & $0.28 {\pm}0.03 $ & $11.6 {\pm}0.4  $ & $11.5 {\pm}0.4  $ & $11.5 {\pm}0.4  $ & $60 {\pm}2  $ \\
                     & $3$ & $0.5  {\pm}0.01 $ & $0.74 {\pm}0.03 $ & $15.5 {\pm}0.4  $ & $15.9 {\pm}0.4  $ & $15.7 {\pm}0.4  $ & $25 {\pm}1  $ \\
                     & $4$ & $1  {\pm}0.02 $ & $0.84 {\pm}0.06 $ & $11.4 {\pm}0.5  $ & $12.3 {\pm}0.5  $ & $11.8 {\pm}0.5  $ & $-8 {\pm}1  $ \\
                     & $5^\star$ & $1.58 {\pm}0.04 $ & $1.8  {\pm}0.2  $ & $6.4  {\pm}0.5  $ & $6.4  {\pm}0.5  $ & $6  {\pm}0.5  $ & $38 {\pm}3  $ \\
                     & $6$ & $2.13 {\pm}0.05 $ & $1.8  {\pm}0.2  $ & $4.7  {\pm}0.6  $ & $6.7  {\pm}0.6  $ & $5.6  {\pm}0.6  $ & $14. {\pm}3  $ \\
                     & $7$ & $3.65 {\pm}0.06 $ & $1.1  {\pm}0.06 $ & $8.8  {\pm}0.4  $ & $10.5 {\pm}0.6  $ & $9.2  {\pm}0.5  $ & $11.2 {\pm}0.8  $ \\
                     & $8$ & $6.8  {\pm}0.1  $ & $0.83 {\pm}0.03 $ & $6.3  {\pm}0.4  $ & $1.4  {\pm}0.7  $ & $2  {\pm}0.2  $ & $-26{\pm}3  $ \\
                     & $9$ & $7.7  {\pm}0.2  $ & $0.39 {\pm}0.02 $ & $0.9  {\pm}0.5  $ & $14.5 {\pm}0.5  $ & $11.3 {\pm}0.4  $ & $-0.2 {\pm}0.3  $ \\
                     & $10$ & $74 {\pm}2  $ & $0.47 {\pm}0.04 $ & $3.1  {\pm}0.07 $ & $6  {\pm}0.1  $ & $4.5  {\pm}0.1  $ & $-0.3 {\pm}0.2  $ \\
\hline
\parbox[t]{2mm}{\multirow{9}{*}{\rotatebox[origin=lb]{90}{\makecell[c]{Z4 \\ $1.9<\Gamma<2$}}}}  & $1$ & $0.3  {\pm}0.03 $ & $0.02 {\pm}0.02 $ & $9.6  {\pm}0.6  $ & $10.4 {\pm}0.7  $ & $9.9  {\pm}0.7  $ & $34 {\pm}2  $ \\
                     & $2$ & $0.65 {\pm}0.02 $ & $0.52 {\pm}0.03 $ & $17.7 {\pm}0.6  $ & $19.9 {\pm}0.7  $ & $18.7 {\pm}0.6  $ & $10.7 {\pm}0.7  $ \\
                     & $3$ & $1.25 {\pm}0.03 $ & $0.86 {\pm}0.09 $ & $9  {\pm}0.7  $ & $11.4 {\pm}0.8  $ & $10 {\pm}0.7  $ & $-7 {\pm}2  $ \\
                     & $4^\star$ & $1.82 {\pm}0.04 $ & $2.7  {\pm}0.4  $ & $4.1  {\pm}0.4  $ & $4  {\pm}0.6  $ & $3.8  {\pm}0.4  $ & $57 {\pm}7  $ \\
                     & $5$ & $2.5  {\pm}0.06 $ & $1.7  {\pm}0.2  $ & $5.2  {\pm}0.4  $ & $8.1  {\pm}0.7  $ & $6.6  {\pm}0.5  $ & $18 {\pm}2  $ \\
                     & $6$ & $4.46 {\pm}0.09 $ & $1.1  {\pm}0.08 $ & $7.7  {\pm}0.4  $ & $12.2 {\pm}0.7  $ & $9.6  {\pm}0.5  $ & $6.3  {\pm}0.6  $ \\
                     & $7$ & $7.5  {\pm}0.1  $ & $1.1  {\pm}0.1  $ & $4.2  {\pm}0.4  $ & $4.8  {\pm}0.8  $ & $3.9  {\pm}0.6  $ & $2  {\pm}2  $ \\
                     & $8$ & $11 {\pm}0.8  $ & $0.44 {\pm}0.05 $ & $0.7  {\pm}0.4  $ & $9.6  {\pm}0.7  $ & $6.6  {\pm}0.5  $ & $-4.4 {\pm}0.5  $ \\
                     & $9$ & $80 {\pm}4  $ & $0.41 {\pm}0.05 $ & $2.8  {\pm}0.1  $ & $6.2  {\pm}0.2  $ & $5.3  {\pm}0.1  $ & $0.08 {\pm}0.09 $ \\
\hline
\parbox[t]{2mm}{\multirow{9}{*}{\rotatebox[origin=lb]{90}{\makecell[c]{Z5 \\ $2<\Gamma<2.1$}}}}  & $1$ & $0.03 {\pm}0.01 $ & $0.1  {\pm}0.1  $ & $2.4  {\pm}0.4  $ & $3.2  {\pm}0.6  $ & $2.7  {\pm}0.4  $ & $(1{\pm}1){\cdot}10^3$ \\
                     & $2$ & $0.38 {\pm}0.04 $ & $0.2  {\pm}0.07 $ & $9.8  {\pm}0.6  $ & $12.1 {\pm}0.8  $ & $10.8 {\pm}0.7  $ & $-27{\pm}5  $ \\
                     & $3$ & $0.66 {\pm}0.03 $ & $0.89 {\pm}0.08 $ & $12.1 {\pm}0.6  $ & $15.2 {\pm}0.8  $ & $13.5 {\pm}0.7  $ & $30 {\pm}3  $ \\
                     & $4$ & $1.18 {\pm}0.03 $ & $1.1  {\pm}0.1  $ & $8  {\pm}0.7  $ & $9.7  {\pm}0.8  $ & $8.5  {\pm}0.7  $ & $-17{\pm}7  $ \\
                     & $5^\star$ & $2.89 {\pm}0.08 $ & $0.96 {\pm}0.07 $ & $8.3  {\pm}0.4  $ & $4  {\pm}1  $ & $5.6  {\pm}0.2  $ & $58 {\pm}3  $ \\
                     & $6$ & $5.2  {\pm}0.2  $ & $1.4  {\pm}0.2  $ & $4.9  {\pm}0.5  $ & $5.7  {\pm}0.7  $ & $4.3  {\pm}0.5  $ & $13 {\pm}3  $ \\
                     & $7$ & $5.7  {\pm}0.2  $ & $0.32 {\pm}0.02 $ & $3  {\pm}1  $ & $18.1 {\pm}0.6  $ & $13.5 {\pm}0.3  $ & $0.2  {\pm}0.3  $ \\
                     & $8$ & $8.5  {\pm}0.3  $ & $1.2  {\pm}0.2  $ & $3.2  {\pm}0.4  $ & $3.4  {\pm}0.7  $ & $3.1  {\pm}0.5  $ & $5  {\pm}3  $ \\
                     & $9$ & $72 {\pm}7  $ & $0.9  {\pm}0.2  $ & $1.6  {\pm}0.2  $ & $4.3  {\pm}0.3  $ & $2.5  {\pm}0.2  $ & $-0.7 {\pm}0.5  $ \\
\hline
\parbox[t]{2mm}{\multirow{10}{*}{\rotatebox[origin=lb]{90}{\makecell[c]{Z6 \\ $2.1<\Gamma<2.2$}}}} & $1$ & $0.021{\pm}0.004$ & $0.05 {\pm}0.06 $ & $2.1  {\pm}0.1  $ & $3.3  {\pm}0.2  $ & $2.6  {\pm}0.1  $ & $-600{\pm}600$ \\
                     & $2$ & $0.77 {\pm}0.07 $ & $0.03 {\pm}0.03 $ & $7.1  {\pm}0.5  $ & $10.7 {\pm}0.8  $ & $8.6  {\pm}0.6  $ & $27 {\pm}1  $ \\
                     & $3$ & $1.22 {\pm}0.04 $ & $0.59 {\pm}0.04 $ & $10.6 {\pm}0.5  $ & $16.5 {\pm}0.7  $ & $13.2 {\pm}0.5  $ & $6.6  {\pm}0.9  $ \\
                     & $4$ & $2.15 {\pm}0.05 $ & $1  {\pm}0.09 $ & $5.6  {\pm}0.4  $ & $9.3  {\pm}0.7  $ & $6.9  {\pm}0.4  $ & $-17.0{\pm}2  $ \\
                     & $5^\star$ & $3.26 {\pm}0.09 $ & $2.3  {\pm}0.3  $ & $3.2  {\pm}0.3  $ & $3.6  {\pm}0.6  $ & $3.3  {\pm}0.4  $ & $26 {\pm}3  $ \\
                     & $6$ & $4.8  {\pm}0.1  $ & $1.5  {\pm}0.2  $ & $4.2  {\pm}0.4  $ & $8.1  {\pm}0.7  $ & $5.1  {\pm}0.4  $ & $15 {\pm}2  $ \\
                     & $7$ & $8.1  {\pm}0.3  $ & $1.4  {\pm}0.2  $ & $3.3  {\pm}0.3  $ & $3.9  {\pm}0.9  $ & $1.1  {\pm}0.4  $ & $40 {\pm}10 $ \\
                     & $8$ & $9.7  {\pm}0.4  $ & $0.5  {\pm}0.04 $ & $0.8  {\pm}0.5  $ & $11.4 {\pm}0.6  $ & $8.7  {\pm}0.4  $ & $-1.3 {\pm}0.4  $ \\
                     & $9$ & $11.8 {\pm}0.3  $ & $1.9  {\pm}0.4  $ & $1.6  {\pm}0.3  $ & $1.6  {\pm}0.6  $ & $1.3  {\pm}0.3  $ & $1  {\pm}4  $ \\
                     & $10$ & $120.0{\pm}10 $ & $0.4  {\pm}0.1  $ & $1.6  {\pm}0.1  $ & $4.5  {\pm}0.3  $ & $3.1  {\pm}0.3  $ & $0.4  {\pm}0.2  $ \\
\hline
\parbox[t]{2mm}{\multirow{9}{*}{\rotatebox[origin=lb]{90}{\makecell[c]{Z7 \\ $2.2<\Gamma<2.3$}}}}  & $1$ & $0.013{\pm}0.002$ & $0.07 {\pm}0.09 $ & $2.9  {\pm}0.2  $ & $4.1  {\pm}0.3  $ & $3.3  {\pm}0.2  $ & $0  {\pm}400$ \\
                     & $2$ & $0.19 {\pm}0.03 $ & $0.03 {\pm}0.04 $ & $2.8  {\pm}0.1  $ & $4.8  {\pm}0.2  $ & $3.6  {\pm}0.2  $ & $10 {\pm}30 $ \\
                     & $3$ & $1.97 {\pm}0.06 $ & $0.34 {\pm}0.02 $ & $9.8  {\pm}0.3  $ & $17.3 {\pm}0.5  $ & $13 {\pm}0.4  $ & $3.1  {\pm}0.7  $ \\
                     & $4$ & $2.67 {\pm}0.06 $ & $0.8  {\pm}0.08 $ & $4.8  {\pm}0.5  $ & $10.1 {\pm}0.9  $ & $6.9  {\pm}0.6  $ & $-7 {\pm}1  $ \\
                     & $5^\star$ & $5  {\pm}0.1  $ & $2.1  {\pm}0.2  $ & $2.9  {\pm}0.3  $ & $4  {\pm}1  $ & $3.5  {\pm}0.5  $ & $12 {\pm}2  $ \\
                     & $6$ & $6.6  {\pm}0.3  $ & $1.2  {\pm}0.1  $ & $2.2  {\pm}0.4  $ & $8.4  {\pm}0.5  $ & $4.6  {\pm}0.4  $ & $5  {\pm}1  $ \\
                     & $7$ & $11.2 {\pm}0.2  $ & $1.6  {\pm}0.2  $ & $1.9  {\pm}0.1  $ & $1.3  {\pm}0.7  $ & $1.5  {\pm}0.2  $ & $18 {\pm}3  $ \\
                     & $8$ & $13.8 {\pm}0.3  $ & $0.86 {\pm}0.06 $ & $0.5  {\pm}0.3  $ & $6.3  {\pm}0.4  $ & $4  {\pm}0.2  $ & $-4.3 {\pm}0.6  $ \\
                     & $9$ & $120.0{\pm}10 $ & $0.4  {\pm}0.1  $ & $0.4  {\pm}0.2  $ & $4.1  {\pm}0.2  $ & $3  {\pm}0.2  $ & $-0.1 {\pm}0.1  $ \\
\hline
\parbox[t]{2mm}{\multirow{8}{*}{\rotatebox[origin=lb]{90}{\makecell[c]{Z8 \\ $2.3<\Gamma<2.4$}}}}  & $1$ & $0.017{\pm}0.002$ & $0.04 {\pm}0.04 $ & $1.93 {\pm}0.09 $ & $3.7  {\pm}0.2  $ & $2.48 {\pm}0.09 $ & $200{\pm}400$ \\
                     & $2$ & $0.41 {\pm}0.05 $ & $0.02 {\pm}0.02 $ & $2.3  {\pm}0.1  $ & $4.9  {\pm}0.3  $ & $3.3  {\pm}0.2  $ & $20 {\pm}10 $ \\
                     & $3$ & $2.3  {\pm}0.1  $ & $0.44 {\pm}0.03 $ & $6.4  {\pm}0.2  $ & $13.3 {\pm}0.5  $ & $9.2  {\pm}0.3  $ & $2.2  {\pm}0.9  $ \\
                     & $4$ & $3.22 {\pm}0.06 $ & $1.01 {\pm}0.08 $ & $3.7  {\pm}0.3  $ & $8.7  {\pm}0.7  $ & $5.6  {\pm}0.5  $ & $-5 {\pm}1  $ \\
                     & $5^\star$ & $6.84 {\pm}0.09 $ & $1.25 {\pm}0.07 $ & $3.2  {\pm}0.1  $ & $4  {\pm}1  $ & $4.3  {\pm}0.3  $ & $1  {\pm}1  $ \\
                     & $6$ & $7.8  {\pm}0.2  $ & $0.95 {\pm}0.07 $ & $0.7  {\pm}0.4  $ & $8.4  {\pm}0.7  $ & $3.2  {\pm}0.4  $ & $12 {\pm}2  $ \\
                     & $7$ & $16.1 {\pm}0.3  $ & $1.07 {\pm}0.08 $ & $1.3  {\pm}0.1  $ & $4.2  {\pm}0.3  $ & $2.8  {\pm}0.1  $ & $-2.6 {\pm}0.6  $ \\
                     & $8$ & $160.0{\pm}30 $ & $0.4  {\pm}0.2  $ & $0.5  {\pm}0.2  $ & $3.2  {\pm}0.3  $ & $2.1  {\pm}0.2  $ & $-0.2 {\pm}0.2  $ \\
\hline
\parbox[t]{2mm}{\multirow{5}{*}{\rotatebox[origin=lb]{90}{\makecell[c]{Z9 \\ $2.4<\Gamma<2.6$}}}}  & $1$ & $0.018{\pm}0.001$ & $0.05 {\pm}0.06 $ & $5  {\pm}0.2  $ & $7.9  {\pm}0.3  $ & $5.6  {\pm}0.2  $ & $-110{\pm}80 $ \\
                     & $2$ & $0.27 {\pm}0.01 $ & $0.03 {\pm}0.03 $ & $4.4  {\pm}0.09 $ & $7.3  {\pm}0.2  $ & $5.2  {\pm}0.1  $ & $40{\pm}6  $ \\
                     & $3$ & $1.6  {\pm}0.1  $ & $0.03 {\pm}0.03 $ & $3.2  {\pm}0.1  $ & $6.7  {\pm}0.2  $ & $4.5  {\pm}0.1  $ & $-4{\pm}2  $ \\
                     & $4$ & $6.3  {\pm}0.1  $ & $0.41 {\pm}0.03 $ & $3  {\pm}0.1  $ & $7.7  {\pm}0.5  $ & $4.9  {\pm}0.2  $ & $-0.8{\pm}0.6  $ \\
                     & $5$ & $13 {\pm}2  $ & $0.08 {\pm}0.04 $ & $0.6  {\pm}0.3  $ & $6.3  {\pm}0.6  $ & $2.8  {\pm}0.3  $ & $2.0{\pm}0.6  $ \\
\hline
\parbox[t]{2mm}{\multirow{5}{*}{\rotatebox[origin=lb]{90}{\makecell[c]{Z10 \\ $\Gamma>2.6$}}}} & $1$ & $0.14 {\pm}0.01 $ & $0.02 {\pm}0.02 $ & $5.9  {\pm}0.1  $ & $8.5  {\pm}0.2  $ & $6.4  {\pm}0.1  $ & $35 {\pm}3  $ \\
                     & $2$ & $0.016{\pm}0.002$ & $0.02 {\pm}0.03 $ & $6.5  {\pm}0.2  $ & $8.9  {\pm}0.3  $ & $6.6  {\pm}0.2  $ & $-150{\pm}20 $ \\
                     & $3$ & $0.79 {\pm}0.05 $ & $0.03 {\pm}0.03 $ & $4.8  {\pm}0.1  $ & $8.1  {\pm}0.3  $ & $5.9  {\pm}0.2  $ & $-6 {\pm}1  $ \\
                     & $4$ & $3.4  {\pm}0.2  $ & $0.04 {\pm}0.03 $ & $2.3  {\pm}0.2  $ & $6.7  {\pm}0.2  $ & $4  {\pm}0.2  $ & $2.8  {\pm}0.4  $ \\
                     & $5$ & $12.9 {\pm}0.5  $ & $0.26 {\pm}0.04 $ & $0.3  {\pm}0.1  $ & $4.1  {\pm}0.2  $ & $2  {\pm}0.1  $ & $-1.9 {\pm}0.7  $ \\
\hline
\end{tabular}
    }
    \tablefoot{Same as \hyperref[tab:plorstats]{Table~\ref{tab:plorstats}} but for the $\tau$-model, where $\Delta t$ corresponds to the time-lag of the Lorentzian.}

\end{table}

\end{appendix}

\end{document}